\documentclass[floatfix,twocolumn,showpacs,preprintnumbers,amsmath,amssymb,pra,superscriptaddress,longbibliography]{revtex4-1}
\usepackage{color}
\usepackage[usenames,dvipsnames,svgnames,table]{xcolor}
\usepackage[colorlinks=true,linkcolor=blue,urlcolor=blue,citecolor=blue]{hyperref}
\usepackage{mathtools}
\usepackage{graphicx}
\usepackage{dcolumn}
\usepackage{array}
\usepackage{lipsum}
\usepackage{bm}
\usepackage{subfigure}
\usepackage{amssymb}
\usepackage{multirow}
\usepackage{tabularx}
\usepackage{amsmath}
\usepackage{braket}
\usepackage{csquotes}
\graphicspath{{plots/}}
 \usepackage{lipsum}
\usepackage{mathrsfs}
\usepackage{MnSymbol}
	
\newcommand{\beq}{\begin{equation}}
\newcommand{\eeq}{\end{equation}}
\newcommand{\bea}{\begin{eqnarray}}
\newcommand{\eea}{\end{eqnarray}}

\begin{document}

\title{Chemical potential of the warm dense electron gas from \emph{ab initio}\\ path integral Monte Carlo simulations}
\author{Tobias Dornheim}
\email{t.dornheim@hzdr.de}
\affiliation{Center for Advanced Systems Understanding (CASUS), D-02826 G\"orlitz, Germany}
\affiliation{Helmholtz-Zentrum Dresden-Rossendorf (HZDR), D-01328 Dresden, Germany}
\author{Michael Bonitz}
\affiliation{Institut f\"ur Theoretische Physik und Astrophysik, Christian-Albrechts-Universit\"at zu Kiel, D-24098 Kiel, Germany}
\affiliation{Kiel Nano, Surface and Interface Science KiNSIS, Christian-Albrechts Universität Kiel, D-24098 Kiel, Germany}
\author{Zhandos~A.~Moldabekov}
\affiliation{Center for Advanced Systems Understanding (CASUS), D-02826 G\"orlitz, Germany}
\affiliation{Helmholtz-Zentrum Dresden-Rossendorf (HZDR), D-01328 Dresden, Germany}
\author{Sebastian Schwalbe}
\affiliation{Center for Advanced Systems Understanding (CASUS), D-02826 G\"orlitz, Germany}
\affiliation{Helmholtz-Zentrum Dresden-Rossendorf (HZDR), D-01328 Dresden, Germany}
\author{Panagiotis Tolias}
\affiliation{Space and Plasma Physics, Royal Institute of Technology (KTH), Stockholm, SE-100 44, Sweden}
\author{Jan Vorberger}
\affiliation{Helmholtz-Zentrum Dresden-Rossendorf (HZDR), D-01328 Dresden, Germany}

\begin{abstract}
We present extensive new \emph{ab initio} path integral Monte Carlo (PIMC) simulation results for the chemical potential of the warm dense uniform electron gas (UEG), spanning a broad range of densities and temperatures. This is achieved by following two independent routes, i) based on the direct estimation of the free energy [Dornheim \emph{et al.}, arXiv:2407.01044] and ii) using a histogram estimator in PIMC simulations with a varying number of particles. We empirically confirm the expected inverse linear dependence of the exchange--correlation (XC) part of the chemical potential on the simulated number of electrons, which allows for a reliable extrapolation to the thermodynamic limit without the necessity for an additional finite-size correction. We find very good agreement (within $\Delta\mu_\textnormal{xc}\lesssim0.5\%$) with the previous parametrization of the XC-free energy by Groth \emph{et al.}~[\emph{Phys.~Rev.~Lett.}~\textbf{119}, 135001 (2017)], which constitutes an important cross validation of current state-of-the-art UEG equations of state. In addition to being interesting in its own right, our study constitutes the basis for the future PIMC based investigation of the chemical potential of real warm dense matter systems starting with hydrogen.
\end{abstract}
\maketitle

\section{Introduction}

Understanding \emph{warm dense matter} (WDM) -- an extreme state characterized by high densities, temperatures, and pressures~\cite{wdm_book,new_POP,drake2018high} -- constitutes a formidable challenge for theory and experiment alike~\cite{bonitz_pop_24}. In nature, such conditions abound in a great variety of astrophysical objects such as giant planet interiors~\cite{Benuzzi_Mounaix_2014,militzer1}, brown dwarfs~\cite{becker}, and white dwarf atmospheres~\cite{SAUMON20221}. In addition, WDM plays an important role in several technological applications~\cite{Kraus2016,Brongersma2015}. A prime example is inertial confinement fusion (ICF)~\cite{Hurricane_RevModPhys_2023,Betti2016}, where both the fusion fuel and its surrounding ablator material have to traverse the WDM regime in a controlled way~\cite{hu_ICF}, while avoiding incipient inhomogeneities and instabilities. Consequently, a rigorous theoretical understanding of WDM is indispensable for the integrated multi-scale modelling of ICF applications which, in turn, is needed for the development of improved designs with an increased fusion yield to make ICF an economically viable future source of clean and abundant energy~\cite{Batani_roadmap}.

In the laboratory, WDM is created using a plethora of different techniques~\cite{falk_wdm} at large research facilities such as the National Ignition Facility (NIF)~\cite{Moses_NIF,Tilo_Nature_2023,MacDonald_POP_2023}, the Linac Coherent Light Source (LCLS)~\cite{LCLS_2016,Glenzer_2016}, and the OMEGA laser facility~\cite{OMEGA,Glenzer_PRL_2007} in the USA, the Laser Megajoule~\cite{Miquel_2016} in France, and the European XFEL~\cite{Tschentscher_2017} in Germany. Here, a major challenge is given by the rigorous diagnostics of the extreme states of matter generated, which is further exacerbated by the ultrafast time scales (typically $10^{-12}-10^{-9}\,$s). In practice, even the most advanced inelastic x-ray scattering methods~\cite{siegfried_review,sheffield2010plasma,Fletcher2015} typically depend on approximate theoretical models for their interpretation~\cite{Gregori_PRE_2003,Tilo_Nature_2023,boehme2023evidence,Fletcher_Frontiers_2022,Sperling_PRL_2015}. Although there has been noteable recent progress both in the model-free interpretation of x-ray Thomson scattering (XRTS) measurements~\cite{Dornheim_T_2022,Dornheim_T_follow_up,dornheim2023xray,dornheim2024modelfreerayleighweightxray,vorberger2023revealing} and in the more rigorous modelling based on \emph{ab initio} simulations~\cite{Schoerner_PRE_2023,Dornheim_Science_2024,Gawne_PRB_2024,Kononov2022}, the value of these experiments for the determination of key properties such as the equation-of-state (EOS) of a given material~\cite{Falk_PRL_2014,Falk_HEDP_2012} has remained limited; an improved theoretical description is therefore strongly needed.

From a theoretical perspective, WDM is more conveniently defined by two dimensionless parameters that are both of the order of unity~\cite{Ott2018,Dornheim_review}: the density parameter $r_s=(3/4\pi n)^{1/3}$ with $n$ the number density, and the degeneracy temperature $\Theta=k_\textnormal{B}T/E_\textnormal{F}$ with $E_\textnormal{F}$ the usual electronic Fermi energy~\cite{quantum_theory}; a third parameter is given by the classical coupling parameter of the ions $\Gamma$ that indicates the ratio of Coulomb interaction to ionic kinetic energy, and is also of the order of or even larger than one. This implies the nontrivial interplay of effects such as Coulomb coupling, strong thermal excitations, quantum degeneracy and diffraction, and, in general, also partial ionization that have to be taken into account holistically. In thermal equilibrium, the gold standard for the modelling of quantum many-body systems such as WDM is given by quantum Monte Carlo (QMC) methods~\cite{anderson2007quantum}, with the \emph{ab initio} path integral Monte Carlo (PIMC) approach~\cite{Berne_JCP_1982,Pollock_PRB_1984,Takahashi_Imada_PIMC_1984} often being the method of choice at finite temperatures. 

In this context, a key system is given by the \emph{uniform electron gas} (UEG)~\cite{quantum_theory,review,loos}, which is the archetypal system of interacting electrons and constitutes the quantum mechanical generalization of the classical one-component plasma. 
In the ground state, QMC results for the UEG~\cite{Ceperley_Alder_PRL_1980,Ortiz_PRB_1994,Ortiz_PRL_1999,moroni,moroni2,Spink_PRB_2013} have been pivotal for the arguably unrivaled success of density functional theory (DFT) in statistical physics, quantum chemistry, material science, and related disciplines~\cite{Jones_RMP_2015}. Over the last decade, these efforts have been extended to WDM conditions~\cite{Brown_PRL_2013,Schoof_PRL_2015,Filinov_PRE_2015,dornheim_prl,groth_prl,Malone_JCP_2015,Malone_PRL_2016,Yilmaz_JCP_2020,Hou_PRB_2022,Joonho_JCP_2021}, resulting in extensive results for a wealth of UEG properties such as its linear~\cite{dornheim_ML,Hou_PRB_2022,groth_jcp,dornheim_electron_liquid,Dornheim_PRL_2020_ESA,Dornheim_PRB_ESA_2021,Dornheim_PRR_2022} and non-linear~\cite{Dornheim_PRL_2020,Dornheim_PRR_2021,Dornheim_JCP_ITCF_2021,Dornheim_JPSJ_2021} density response, its structural properties~\cite{Dornheim_CPP_2017,dornheim_prl,Brown_PRL_2013}, its momentum distribution function~\cite{Hunger_PRE_2021,Dornheim_PRB_nk_2021,Dornheim_PRE_2021,Militzer_Pollock_PRL_2002,MILITZER201913}, its virial coefficients~\cite{Dornheim_HEDP_2022,Roepke_PRE_2024}, its imaginary-time structure~\cite{Dornheim_MRE_2023,Dornheim_PTR_2023,Dornheim_moments_2023}, and even its dynamic structure factor~\cite{dornheim_dynamic,dynamic_folgepaper,Dornheim_Nature_2022,Filinov_PRB_2023} and dynamic Matsubara density response~\cite{Tolias_JCP_2024,Dornheim_PRB_2024,Dornheim_EPL_2024,dornheim2024shortwavelengthlimitdynamic}.
Of particular value are the accurate parametrizations of its exchange--correlation (XC) free energy $F_\textnormal{xc}$~\cite{groth_prl,review,ksdt,status} that continuously cover the entire relevant WDM phase diagram region. These XC-functionals can be used directly as input for thermal DFT simulations~\cite{Mermin_DFT_1965} of real WDM applications on the level of the local density approximation~\cite{karasiev_importance,Sjostrom_PRB_2014,kushal}, and as the basis for the construction of more advanced functionals~\cite{Sjostrom_PRB_2014,Karasiev_PRL_2018,Karasiev_PRB_2022,kozlowski2023generalized}.

In practice, the construction of the two most important $F_\textnormal{xc}$ parametrizations -- the functional by Groth \emph{et al.}~\cite{groth_prl} (GDSMFB) and the corrected functional by Karasiev \emph{et al.}~\cite{Karasiev_PRL_2018,ksdt} (corrKSDT) -- is based on PIMC results for the interaction energy $W$, kinetic energy $K$, or internal energy $E=W+K$ (or a combination of these three) as well as on thermodynamic relations such as the well-known adiabatic connection formula~\cite{IIT}. This was necessary as PIMC by itself does not give one direct access to the partition function $Z(N,V,\beta)$, with $N$ the total particle number, $V=L^3$  the volume of the cubic simulation cell, and $\beta=1/k_\textnormal{B}T$ the inverse temperature in energy units, thus precluding direct access to the free energy
\begin{eqnarray}\label{eq:F_from_Z}
    F(N,V,\beta)=-\frac{1}{\beta}\textnormal{log}Z(N,V,\beta)\ .
\end{eqnarray}
Very recently, four of us have changed this rather unfortunate situation by introducing simulations in the extended $\eta$-ensemble~\cite{dornheim2024directfreeenergycalculation}. This new scheme connects a non-interacting and uniform Bose reference system, for which the canonical partition function and free energy can be readily estimated from a well-known recursion relation~\cite{krauth2006statistical,Bargathi_PRR_2020,Zhou_2018}, with the interacting (and, in general, potentially non-uniform) system of interest and, in this way, allows for the direct estimation of the free energy from PIMC simulations for a single density--temperature combination, i.e., without thermodynamic integration. The subsequent extensive application of this idea to the warm dense UEG~\cite{Dornheim_F_Follow_up} has fully substantiated the high quality of the GDSMFB parametrization, which constitutes a valuable independent cross-check of this key property.

In the present work, we consider another important thermodynamic property: the chemical potential $\mu(N,V,\beta)$ that describes the change in the free energy upon adding an additional particle to the system.
In general, the accurate knowledge of $\mu$ is crucial for understanding the stability conditions of complex systems. Indeed, the chemical potential governs the stability of multiphase systems (it yields the phase coexistence line), the chemical stability (equilibrium between reaction products, mass action law) and the stable density profile in an external field. Moreover, the chemical potential provides some insight into the electronic properties. Since it is related to the energy needed to add particles to the system, $\delta W_{add}=\mu dN$, it contains information about the binding energy and the energy of the highest occupied orbital (HOMO) in a quantum system either in the ground state or in the presence of a band gap that is larger than the thermal energy. In finite systems, such as nuclei, Coulomb clusters \cite{tsuruta_pra_93,ludwig_pre_5,bonitz_rpp_10}, ions in traps \cite{dubin_rmp_99} or artificial atoms \cite{ashoori_nat_96,bedanov_prb_94}, $\mu$ is directly related to the addition energy and to the electronic structure (shell configuration, magic clusters etc.). In a macroscopic Fermi system in the ground state, it coincides with the Fermi energy. We underline that this applies not only to ideal models, but also to interacting systems, where the Fermi energy is generally not known. Therefore, a first-principles approach to the chemical potential will be extremely valuable, in particular for warm dense matter. 

In classical systems, the chemical potential has been computed from numerical simulations using free energy perturbation methods~\cite{Bennet1976}, Widom insertion methods~\cite{Kofke1997}, expanded ensemble methods~\cite{Attard1993}, histogram-reweighting methods~\cite{Shing1982} or spatially resolved thermodynamic integration methods~\cite{Heidari2018}. Most widespread are the free energy perturbation methods that are based on the Zwanzig identity~\cite{Zwanzig1954} and the closely related Widom insertion methods that are based on the potential distribution theorem~\cite{Widom1982,Widom_JCP_2022}. Dedicated umbrella sampling techniques and multistage variants of these methods have been developed with the objective of increasing the sampling efficiency for dense liquids~\cite{Kofke1997,Heidari2018}. However, the theoretical backbone of most of these methods (the Zwanzig identity or the potential distribution theorem) cannot be straightforwardly extended to quantum systems, (i) being based on the classical factorization $\exp{[-\beta{H}(\boldsymbol{r},\boldsymbol{p})]}=\exp{[-\beta{K}(\boldsymbol{p})]}\exp{[-\beta{W}(\boldsymbol{r})]}$ which does not generally hold, since the kinetic and potential energy operators do not commute, and  (ii) not considering the proper Fermi-Dirac (or Bose-Einstein) statistics~\cite{Herdman_PRB_2014}. Here, we use exact PIMC simulations to compute $\mu$ in two independent ways, namely by computing the difference in the free energy [Eq.~(\ref{eq:Mu_from_F})] using our recent $\eta$-ensemble approach and also by estimating the corresponding ratio of $N$- and $N+1-$particle partition functions from a single simulation in which the number of particles is allowed to vary. We find perfect agreement between these routes, as it is expected.

A great practical advantage of the chemical potential $\mu(N,V,\beta)$ concerns its known scaling with respect to the number of electrons $N$~\cite{Herdman_PRB_2014,Siepmann_1992} towards the thermodynamic limit (TDL), i.e., the limit of $N,V\to\infty$ where $n=N/V$ is being kept constant. This allows us to reliably extrapolate our PIMC results to the TDL without the need for the usual semi-analytical finite-size corrections~\cite{Chiesa_PRL_2006,Drummond_PRB_2008,Brown_PRL_2013,dornheim_prl,Dornheim_JCP_2021,uegpy,Holzmann_PRB_2016} that have been used for the construction of the GDSMFB and corrKSDT parametrizations, but also for the finite-size correction of the new free energy results in Refs.~\cite{dornheim2024directfreeenergycalculation,Dornheim_F_Follow_up}. The present estimation of $\mu(N,V,\beta)$ [and its XC-contribution $\mu_\textnormal{xc}(N,V,\beta)]$ thus constitutes a completely independent route towards the EOS of a given system, and thus allows us to further test previous results over a broad range of densities and temperatures.
In addition to its considerable value for the further understanding of the archetypal UEG, the present study will serve as a template for the future PIMC based investigation of the chemical potential of real WDM systems such as hydrogen~\cite{filinov2023equation,bonitz_pop_24,Dornheim_JCP_2024,Bohme_PRL_2022,Militzer_PRE_2001,Militzer_PRE_2021} and beryllium~\cite{Dornheim_Science_2024}.

The paper is organized as follows: In Sec.~\ref{sec:theory}, we give a concise overview of the required theoretical background, including an introduction to the chemical potential (Sec.~\ref{sec:mu}), to the direct PIMC estimation of the free energy via the extended $\eta$-ensemble (Sec.~\ref{sec:eta_ensemble}), to the estimation of the canonical properties of the ideal Bose gas and ideal Fermi gas (Sec.~\ref{sec:ideal}), to the decomposition of the total chemical potential into an ideal, a bosonic interaction, and a quantum statistical contribution (Sec.~\ref{sec:final}), to the alternative histogram estimator route (Sec.~\ref{sec:hist}), and to the estimation of $\mu_\textnormal{xc}$ based on parametrizations of $f_\textnormal{xc}(r_s,\Theta)$~(\ref{sec:EOS}).
All simulation results are presented in Sec.~\ref{sec:results}, starting with the verification and comparison of the independent PIMC estimators~(\ref{sec:verification}) and an exploration of the required extrapolation to the thermodynamic limit (Sec.~\ref{sec:extrapolation}); the bulk of our results is presented in Sec.~\ref{sec:parametrization}, where we compare the present data to the previous parametrization by Groth \emph{et al.}~\cite{groth_prl}.
The paper is concluded by a summary and outlook in Sec.~\ref{sec:outlook}.

\section{Theory\label{sec:theory}}

We assume Hartree atomic units throughout this work unless indicated otherwise. Detailed introductions to the PIMC method~\cite{cep,boninsegni2}, its application to the UEG~\cite{review,Dornheim_permutation_cycles}, and the associated fermion sign problem~\cite{dornheim_sign_problem} have been presented in the literature.

\subsection{Chemical potential\label{sec:mu}}

In the canonical ensemble, which is defined by a constant number of particles $N$, volume $V=L^3$ and inverse temperature $\beta=1/T$, the chemical potential $\mu(N,V,\beta)$ is given by the change in the free energy  associated with the addition of a particle, $\{N,V,\beta\}\to\{N+1,V,\beta\}$,
\begin{eqnarray}\label{eq:Mu_from_F}
    \mu(N,V,\beta) \equiv F(N+1,V,\beta) - F(N,V,\beta)\ ,
\end{eqnarray}
with the free energy related to the canonical partition function $Z(N,V,\beta)$ by the usual relation, Eq.~(\ref{eq:F_from_Z}). While both $F(N,V,\beta)$ and $Z(N,V,\beta)$ cannot be directly estimated within a traditional PIMC simulation, one can obtain both by connecting the interacting (and, in general, also inhomogeneous) system of interest with a non-interacting reference system for which both $F$ and $Z$ are known~\cite{dornheim2024directfreeenergycalculation,Dornheim_F_Follow_up}, see Sec.~\ref{sec:eta_ensemble} below.

For simulations with periodic boundary conditions,
it has become a standard practice to use the Ewald pair interaction potential, which constitutes the corresponding solution of Poisson's equation~\cite{Fraser_PRB_1996}. In this situation, the definition of Eq.~(\ref{eq:Mu_from_F}) is somewhat ambiguous. Specifically, one might either add a Coulomb interacting electron (i.e., without the infinite periodic array of images), or an additional Ewald interacting electron with its own neutralizing background. Bakhshandeh and Levin~\cite{Widom_JCP_2022} have shown that, for classical systems, while both approaches converge to the same TDL, the latter exhibits a more favorable converge behavior. In the present quantum case, it is important that the added electron is indistinguishable from the other electrons (of the same spin orientation), which makes the neutralized Ewald potential the appropriate choice.

\subsection{Free energy and the $\eta$-ensemble\label{sec:eta_ensemble}}

To compute the free energy from PIMC simulations, we follow the approach introduced in Refs.~\cite{dornheim2024directfreeenergycalculation,Dornheim_F_Follow_up} and consider the modified Hamiltonian
\begin{eqnarray}\label{eq:Hamiltonian_eta}
    \hat{H}_\eta = \hat{K} + \eta \hat{W}\ ,
\end{eqnarray}
where $\hat{K}$ and $\hat{W}$ denote the operators of kinetic and potential energy, respectively. Here $\eta\in[0,1]$ denotes a free parameter, with $\eta=0$ describing a non-interacting reference system and $\eta=1$ corresponding to the non-ideal system of interest.
By connecting these two limits (with $N_\eta\geq1$ intermediate steps), one can access the free energy of the interacting Fermi system via~\cite{dornheim2024directfreeenergycalculation}
\begin{eqnarray}\label{eq:F_eta_final}
    F_\textnormal{F}(N,V,\beta) &=& F_{\textnormal{B},0}(N,V,\beta)\\\nonumber & & - \frac{1}{\beta}\left\{ \sum_{i=1}^{N_\eta}\textnormal{log}\left( \frac{r(\eta_i,\eta_{i+1})}{c_{\eta_i}} \right) + \textnormal{log}\left(S\right)
    \right\}\ ,
\end{eqnarray}
with $F_{\textnormal{B},0}(N,V,\beta)$ the free energy of an ideal Bose gas that can be computed from a recursion relation for the corresponding canonical partition function, cf.~Eq.~(\ref{eq:Z_ideal_Bose_Fermi}) below.
The second term on the RHS~of Eq.~(\ref{eq:F_eta_final}) connects the ideal and non-ideal Bose systems, with $r(\eta_i,\eta_{i+1})$ being the ratio of configurations with two different $\eta$-values, and with $c_{\eta_i}$ being a free algorithmic parameter; see Ref.~\cite{Dornheim_F_Follow_up} for an extensive discussion. The final term is given by the \emph{average sign}
\begin{eqnarray}\label{eq:sign}
    S = \frac{ Z_\textnormal{F}(N,V,\beta) }{Z_\textnormal{B}(N,V,\beta)}\ ,
\end{eqnarray}
which constitutes an important measure for the degree of cancellation between positive and negative contributions in a PIMC simulation with a sign problem~\cite{dornheim_sign_problem}, and which, in the context of the present work, constitutes a quantum statistics correction connecting the interacting Bose and Fermi systems.

\begin{figure}\centering
\hspace*{-0.017\textwidth}\includegraphics[width=0.465\textwidth]{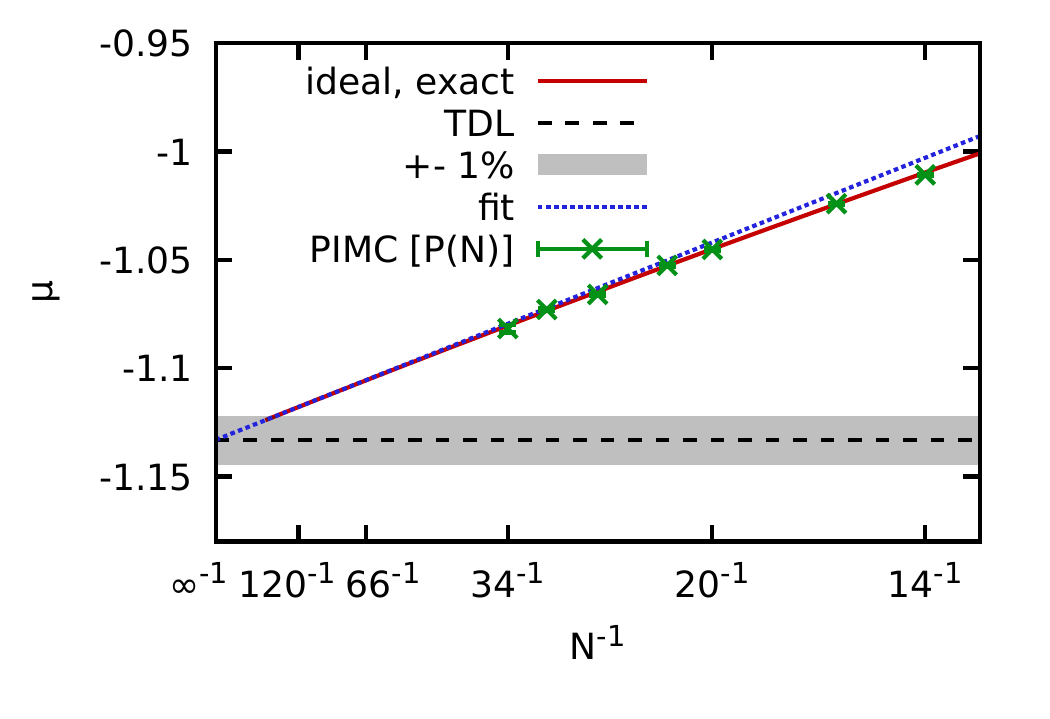}\\\vspace*{-1.35cm}
\includegraphics[width=0.45\textwidth]{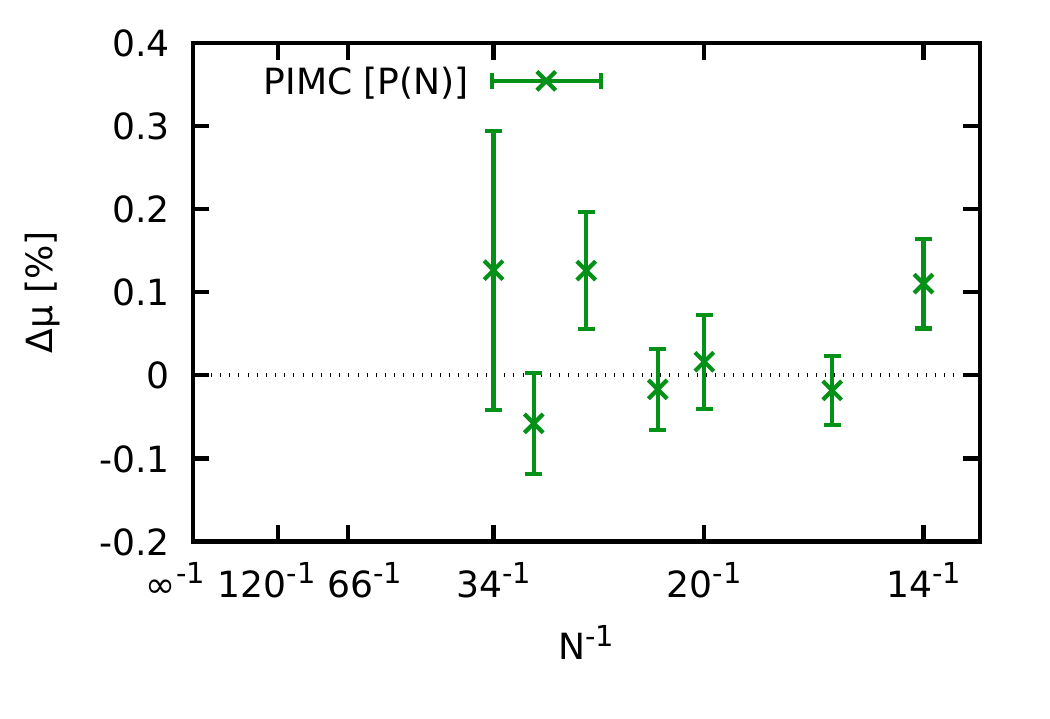}
\caption{\label{fig:ideal_theta2} Top panel: The dependence of the chemical potential of the unpolarized ideal Fermi gas at $r_s=2$ and $\Theta=2$ on the total particle number $N$. Solid red: exact canonical result [see Eq.~(\ref{eq:mu_ideal_spin})]; dashed black: exact thermodynamic limit (TDL); dotted blue: an empirical linear fit for $N\geq100$; green crosses: results of PIMC simulations based on the histogram estimator [see Eq.~(\ref{eq:histrogram_estimator})]. Bottom panel: the relative deviation between the PIMC result and the exact canonical result.}
\end{figure}

\subsection{Ideal Fermi and Bose gas\label{sec:ideal}}

The final missing ingredient is given by $F_{\textnormal{B},0}(N,V,\beta)$. For this purpose, we use the well-known recursion relation of the canonical partition function of ideal quantum many-body systems~\cite{Bargathi_PRR_2020,krauth2006statistical};
a convenient compact form for $N^\uparrow$ spin-polarized bosons or fermions is given by~\cite{Zhou_2018}
\begin{eqnarray}\label{eq:Z_ideal_Bose_Fermi}
    Z_{\textnormal{F/B},0}^\uparrow(N^\uparrow,V,\beta) = \frac{1}{N^\uparrow!} \textnormal{det}\left( \mathbf{Z}^\uparrow_{\textnormal{F/B},0}\right)\ ,
\end{eqnarray}
with the $N^\uparrow\times N^\uparrow$ matrix $\mathbf{Z}^\uparrow_{\textnormal{F/B},0}$ being defined as
\begin{eqnarray}\label{eq:get_off_my_case}
    \mathbf{Z}^\uparrow_{\textnormal{F/B},0}(a,b) = \begin{cases}
    \pm a, & \text{if } a=b+1\\
    Z(1,V,\beta (1+b-a)), & \text{if } a \leq b\\
    0,              & \text{otherwise}\ ,
\end{cases}
\end{eqnarray}
with $1\leq a,b \leq N^\uparrow$; note that $+a$ and $-a$ in the first line of Eq.~(\ref{eq:get_off_my_case}) correspond to Fermi-Dirac and Bose-Einstein statistics, respectively. 
In the noninteracting case, the partition functions of the individual spin-components factorize (this holds even in the spin-polarized case with $N=N^\uparrow$ and $N^\downarrow=0$, as it is $Z^\downarrow=1$), and the full partition function is given by
\begin{eqnarray}
    Z_{\textnormal{F/B},0}(N,V,\beta) =  Z_{\textnormal{F/B},0}^\uparrow(N^\uparrow,V,\beta)\  Z_{\textnormal{F/B},0}^\downarrow(N^\downarrow,V,\beta)\ . \
\end{eqnarray}
It is easy to see that the chemical potential can be expressed as the ratio of partition functions~\cite{Herdman_PRB_2014,hansen2013theory}
\begin{eqnarray}\label{eq:Mu_from_Z}
    \mu(N,V,\beta) = -\frac{1}{\beta} \textnormal{log}\left\{\frac{Z(N+1,V,\beta)}{Z(N,V,\beta)}\right\}\ ;
\end{eqnarray}
the ideal chemical potential that corresponds to the addition of a spin-up particle is thus given by
\begin{eqnarray}\label{eq:mu_ideal_spin}
    \mu^\uparrow(N,V,\beta) = - \frac{1}{\beta} \textnormal{log}\left\{
\frac{Z_{\textnormal{F/B},0}^\uparrow(N^\uparrow+1,V,\beta)}{Z_{\textnormal{F/B},0}^\uparrow(N^\uparrow,V,\beta)}
    \right\}
\end{eqnarray}
as all spin-down contributions trivially cancel. Naturally, Eq.~(\ref{eq:mu_ideal_spin}) does not generalize to interacting systems.

In the top panel of Fig.~\ref{fig:ideal_theta2}, we show the canonical chemical potential of the unpolarized ideal Fermi gas at $r_s=2$ and $\Theta=2$ as a function of the total number of particles. Specifically, the solid red line corresponds to Eq.~(\ref{eq:mu_ideal_spin}) and the dashed black line depicts the thermodynamic limit, i.e., the limit of $N\to\infty$ as the density is being kept constant. We note that the evaluation of the determinant in Eq.~(\ref{eq:Z_ideal_Bose_Fermi}) becomes unstable for $N\gtrsim200$ at these conditions. The dotted blue line shows an empirical linear fit to the canonical results for $N\geq100$ and leads to good agreement with the correct thermodynamic limit. The PIMC results for $\mu$ that are also shown in Fig.~\ref{fig:ideal_theta2} are discussed in Sec.~\ref{sec:hist} below.

\subsection{Decomposition of the chemical potential\label{sec:final}}

Taken together, we can express the chemical potential of the non-ideal Fermi system of interest as
\begin{widetext}
\begin{eqnarray}\label{eq:mu_final}
    \mu_\textnormal{F}(N,V,\beta) = \mu_{\textnormal{B}0}(N,V,\beta)\,
   \underbrace{ - \frac{1}{\beta}\left(
\sum_{i=1}^{N_{\eta,N+1}}\textnormal{log}\left( \frac{r_{N+1}(\eta_i,\eta_{i+1}) }{c^{N+1}_{\eta_i}} \right)
-\sum_{i=1}^{N_{\eta,N}}\textnormal{log}\left( \frac{r_N(\eta_i,\eta_{i+1}) }{c^N_{\eta_i}} \right)
    \right)}_{\Delta\mu_{\textnormal{B}0,\textnormal{B}}(N,V,\beta)}
  \ \   \underbrace{- \frac{1}{\beta}\textnormal{log}\left(\frac{S(N+1,V,\beta)}{S(N,V,\beta)}\right)}_{\Delta\mu_{\textnormal{B},\textnormal{F}}(N,V,\beta) }\ ,\quad 
\end{eqnarray}
\end{widetext}
which consists of three individual contributions that are defined in analogy to the decomposition of the total free energy introduced in Ref.~\cite{Dornheim_F_Follow_up}:
\begin{eqnarray}\label{eq:Pam_Bondi}
\mu_{\textnormal{B}0}(N,V,\beta) &=& F_{\textnormal{B}0}(N+1,V,\beta) - F_{\textnormal{B}0}(N,V,\beta) \\ 
    \nonumber \Delta\mu_{\textnormal{B}0,\textnormal{B}}(N,V,\beta) &=& \Delta F_{\textnormal{B}0,\textnormal{B}}(N+1,V,\beta) - \Delta F_{\textnormal{B}0,\textnormal{B}}(N,V,\beta) \\
    \Delta\mu_{\textnormal{B},\textnormal{F}}(N,V,\beta) &=& \Delta F_{\textnormal{B},\textnormal{F}}(N+1,V,\beta)  -  \Delta F_{\textnormal{B},\textnormal{F}}(N,V,\beta)\ . \nonumber 
\end{eqnarray}
In a nutshell, the full chemical potential is given by the sum of the chemical potential of the ideal Bose system $\mu_{\textnormal{B}0}(N,V,\beta)$, the bosonic interaction contribution $\Delta\mu_{\textnormal{B}0,\textnormal{B}}(N,V,\beta)$, and the quantum statistical contribution $\Delta\mu_{\textnormal{B},\textnormal{F}}(N,V,\beta)$.

\subsection{Chemical potential from particle number histograms\label{sec:hist}}

As an alternative route towards the chemical potential of interacting quantum many-body systems, we can estimate the corresponding ratio of canonical partition functions [Eq.~(\ref{eq:Mu_from_Z})] from PIMC simulations where the particle number is allowed to vary. For example, one might consider the grand-canonical partition function
\begin{eqnarray}\label{eq:Z_GC}
    Z_\textnormal{GC}(\mu_\textnormal{GC},V,\beta) = \sum_{N=0}^\infty e^{\beta \mu_\textnormal{GC} N} Z(N,V,\beta) \ ,
\end{eqnarray}
where the grand-canonical chemical potential $\mu_\textnormal{GC}$ is a free input parameter. The canonical expectation value of $\mu(N,V,\beta)$ then immediately follows as~\cite{Herdman_PRB_2014}
\begin{eqnarray}\label{eq:mu_GC}
    \mu(N,V,\beta) = \mu_\textnormal{GC} - \frac{1}{\beta}\textnormal{log}\left(
\frac{P(N+1)}{P(N)}
    \right)\ ,
\end{eqnarray}
 with $P(N)$ being the histogram of particle numbers, i.e., the expectation value of the operator
\begin{eqnarray}
   \delta_{N}(\mathbf{X}) = \begin{cases}
    1, & \text{if } N = N(\mathbf{X})\\
    0,              & \text{otherwise}\ ;
\end{cases}
\end{eqnarray}
in the grand-canonical simulation, with $N(\mathbf{X})$ being the particle number of the PIMC configuration $\mathbf{X}$; hence,
\begin{eqnarray}\label{eq:P}
    P(N) = \braket{\hat{\delta}_N}_\textnormal{GC}\ .
\end{eqnarray}
We note that, for a fermionic PIMC simulation that is subject to the fermion sign problem~\cite{Dornheim_JPA_2021}, 
the grand-canonical expectation value in Eq.~(\ref{eq:P}) is given by 
\begin{eqnarray}
    \braket{\hat{\delta}_N}_\textnormal{GC} = \frac{\braket{\hat{\delta}_N\hat{S}}_\textnormal{GC}'}{\braket{\hat{S}}_\textnormal{GC}'}\ ,
\end{eqnarray}
with $\braket{\hat{S}}_\textnormal{GC}'$ the grand-canonical average sign~\cite{Dornheim_JPA_2021}, and where the notation $\braket{\dots}_\textnormal{GC}'$ implies the usual expectation value with respect to the bosonic absolute weights~\cite{dornheim_sign_problem,Dornheim_JPA_2021}.

On one hand, the histogram approach given in Eq.~(\ref{eq:mu_GC}) has the advantage that one can get $\mu(N,V,\beta)$ from a single PIMC simulation, and without the need for any external input for the noninteracting Bose- or Fermi-system. On the other hand, a substantial amount of simulation time might be wasted for particle numbers $N'>N+1$ or $N'<N$ that are not relevant for the evaluation of Eq.~(\ref{eq:mu_GC}).
To remedy this issue, we generalize Eq.~(\ref{eq:Z_GC}) as
\begin{eqnarray}\label{eq:Z_GC_general}
    \tilde{Z}_\textnormal{GC}(\mu_\textnormal{GC},V,\beta) = \sum_{N=0}^\infty W(N) Z(N,V,\beta) \ ,
\end{eqnarray}
leading to
\begin{eqnarray}\label{eq:mu_GC_general}
    \mu(N,V,\beta) = -\frac{1}{\beta} \textnormal{log}\left(
\frac{W(N)\braket{\hat{S}\hat{\delta}_{N+1}}_\textnormal{GC}'}{W(N+1)\braket{\hat{S}\hat{\delta}_{N}}_\textnormal{GC}'}
    \right)\ .
\end{eqnarray}
A convenient empirical choice for the generalized weight function $W(N)$ is given by 
\begin{eqnarray}\label{eq:generalized_weight}
    W(N') = e^{\beta\mu_\textnormal{GC}N'} e^{-(N'-N)^2/\sigma_N^2}\ ,
\end{eqnarray}
with $N$ being the target number of particles; the variance $\sigma_N$ controls the amount of irrelevant particle numbers in the PIMC simulation, with $\sigma_N=0.6$ being a good choice. This leads to our final result for the canonical chemical potential,
\begin{eqnarray}\label{eq:histrogram_estimator}
    \mu(N,V,\beta) = \mu_\textnormal{GC} - \frac{1}{\beta}\left\{
\textnormal{log}\left( \frac{\braket{\hat{S}\hat{\delta}_{N+1}}_\textnormal{GC}'}{\braket{\hat{S}\hat{\delta}_{N}}_\textnormal{GC}'} \right) + \frac{1}{2\sigma_N^2}
    \right\}\ .
\end{eqnarray}
Without loss of generality, we employ the full grand-canonical worm algorithm~\cite{boninsegni1,boninsegni2} to update the $N^\uparrow$ spin-up electrons, which are allowed to vary with respect to the number of particles. In contrast, we use a canonical adaption of the worm algorithm~\cite{Dornheim_PRB_nk_2021} to update the $N^\downarrow$ spin-down electrons, for which the number of particles is being kept fixed.

In practice, we find that the histogram approach constitutes a more convenient and efficient route towards $\mu(N,V,\beta)$ and we use it for the bulk of the warm dense UEG results that are presented below. The direct free energy approach is less efficient, but it constitutes a valuable and independent cross check of our implementation.

\begin{figure}\centering
\includegraphics[width=0.45\textwidth]{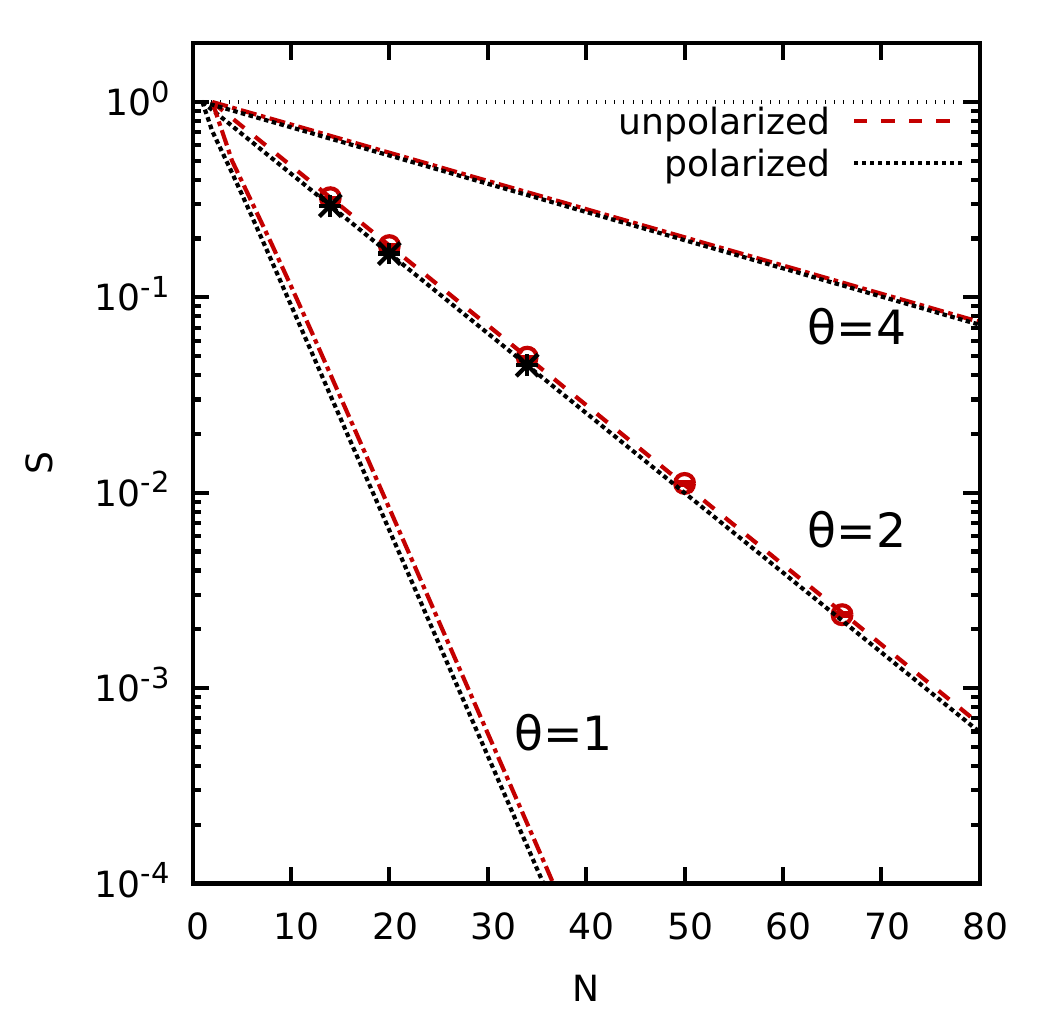}
\caption{\label{fig:ideal_sign} The average sign $S$ as a function of the total particle number $N$ for the fully polarized (dotted black) and unpolarized (dashed red) ideal Fermi gas for $\Theta=4$, $\Theta=2$, and $\Theta=1$. The symbols depict corresponding PIMC results.
}
\end{figure} 

The green crosses in the top panel of Fig.~\ref{fig:ideal_theta2} show PIMC results that have been computed via Eq.~(\ref{eq:histrogram_estimator}) for $\mu_\textnormal{GC}=-0.1$. They are in excellent agreement with the exact reference curve within the given statistical confidence interval; this can be seen particularly well in the bottom panel of Fig.~\ref{fig:ideal_theta2}, where we show the relative deviation of the PIMC data from the exact result. We find increasing error bars upon increasing the system size, which is a direct consequence of the inherent cancellation of positive and negative terms due to the fermion sign problem~\cite{dornheim_sign_problem,Dornheim_JPA_2021}. In Fig.~\ref{fig:ideal_sign}, we show the average sign [Eq.~(\ref{eq:sign})] of the ideal Fermi gas as a function of the system size $N=N^\uparrow + N^\downarrow$ both for the unpolarized (dashed red) and spin-polarized (dotted black) cases for $\Theta=4$, $\Theta=2$, $\Theta=1$. In all three cases, we find the expected exponential decrease with $N$~\cite{troyer,dornheim_sign_problem}, which illustrates the challenges associated with the PIMC simulation of quantum degenerate Fermi systems. Fortunately, the sign problem is considerably less severe for interacting systems, since Coulomb repulsion effectively suppresses the formation of permutation cycles~\cite{Dornheim_permutation_cycles}; this holds both for the extended $\eta$-ensemble and the histogram estimator. The red circles and black stars in Fig.~\ref{fig:ideal_sign} show PIMC results for both spin polarizations for $\Theta=2$ and are in perfect agreement with the exact, semi-analytical result, as expected.

\subsection{Chemical potential from the UEG equation-of-state\label{sec:EOS}}

To utilize available parametrizations of the free energy in terms of $r_s$ and $\Theta$, we need to perform a derivative of the free energy with respect to the density under constant temperature and volume. We define the free energy per particle $f=F/N$, leading to
\begin{equation}
\mu=\frac{\partial F}{\partial N}\Bigg|_{T,V}=f+n\frac{\partial f}{\partial n}\Bigg|_{T,V}\,.
\end{equation}
The derivative of the free energy with respect to the density at constant temperature can be obtained from a parametrization in $r_s$ and $\Theta$ by considering the derivative of a function f(x) in the direction given by vector ${\bf v}$ $\nabla_v f(x)=\nabla f \cdot {\bf v}$.  Using the connection between $r_s$ and $\theta$ for constant temperature $2\theta/r_s\,dr_s=d\Theta$, it follows
\begin{equation}
    \frac{\partial f}{\partial n} = \nabla f \cdot {\bf v}
    =\left[
    \frac{\partial f}{\partial r_s}\Bigg|_{\Theta}+
    \frac{2\Theta}{r_s}\frac{\partial f}{\partial \Theta}\Bigg|_{r_s}
    \right]
    \frac{dr_s}{dn}\,.
\end{equation}
The chemical potential is then
\begin{equation}\label{eq:mu_from_rs_theta}
\mu(r_s,\Theta)=f(r_s,\Theta)-
    \frac{r_s}{3}\frac{\partial f(r_s,\Theta)}{\partial r_s}\Bigg|_{\Theta}-
    \frac{2\Theta}{3}\frac{\partial f(r_s,\Theta)}{\partial \Theta}\Bigg|_{r_s}
    \,.
\end{equation}
Naturally, the same relation holds between the respective XC-contributions $\mu_\textnormal{xc}$ and $f_\textnormal{xc}$.




\section{Results\label{sec:results}}

\begin{figure*}\centering
\includegraphics[width=0.45\textwidth]{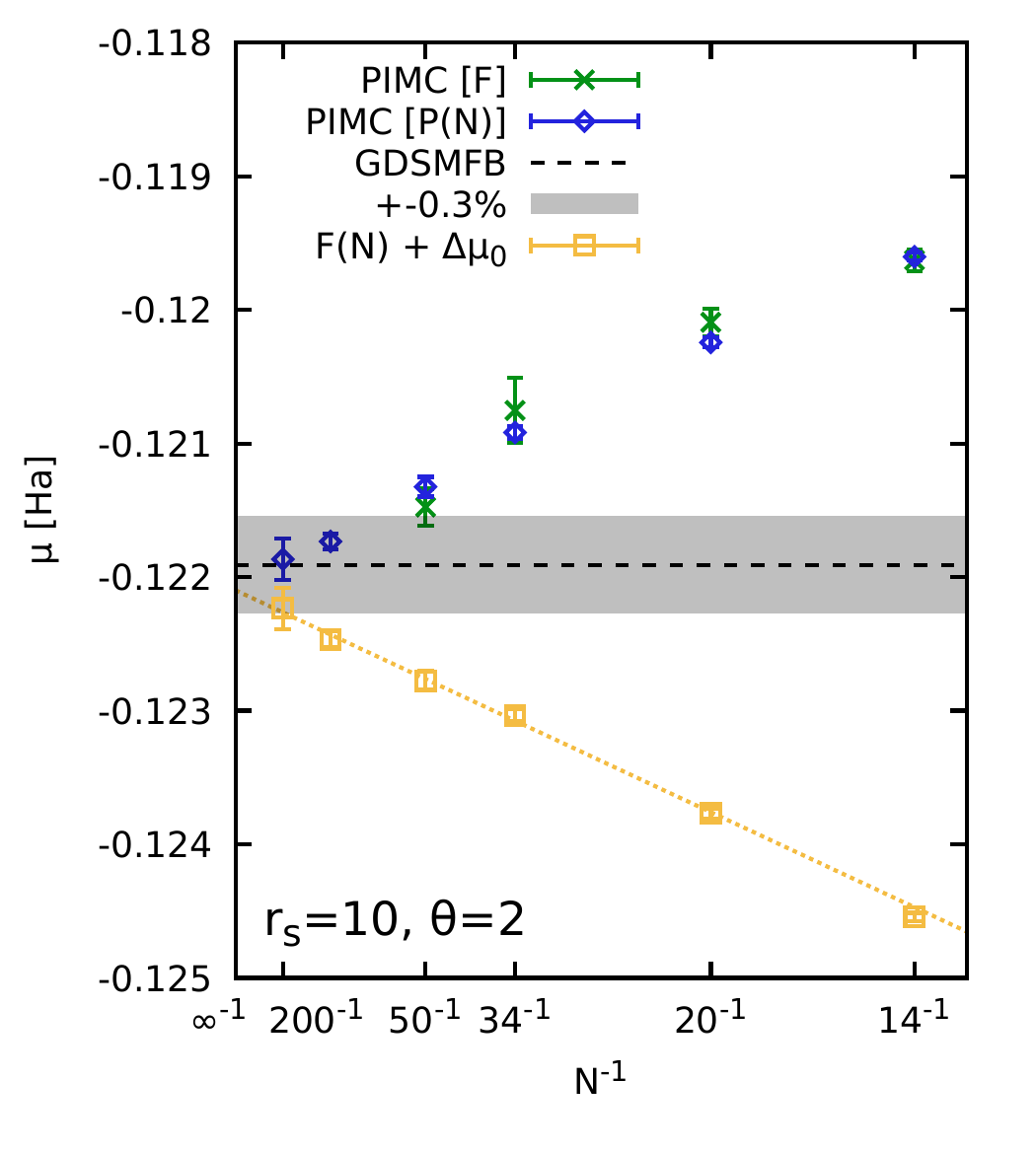}\includegraphics[width=0.45\textwidth]{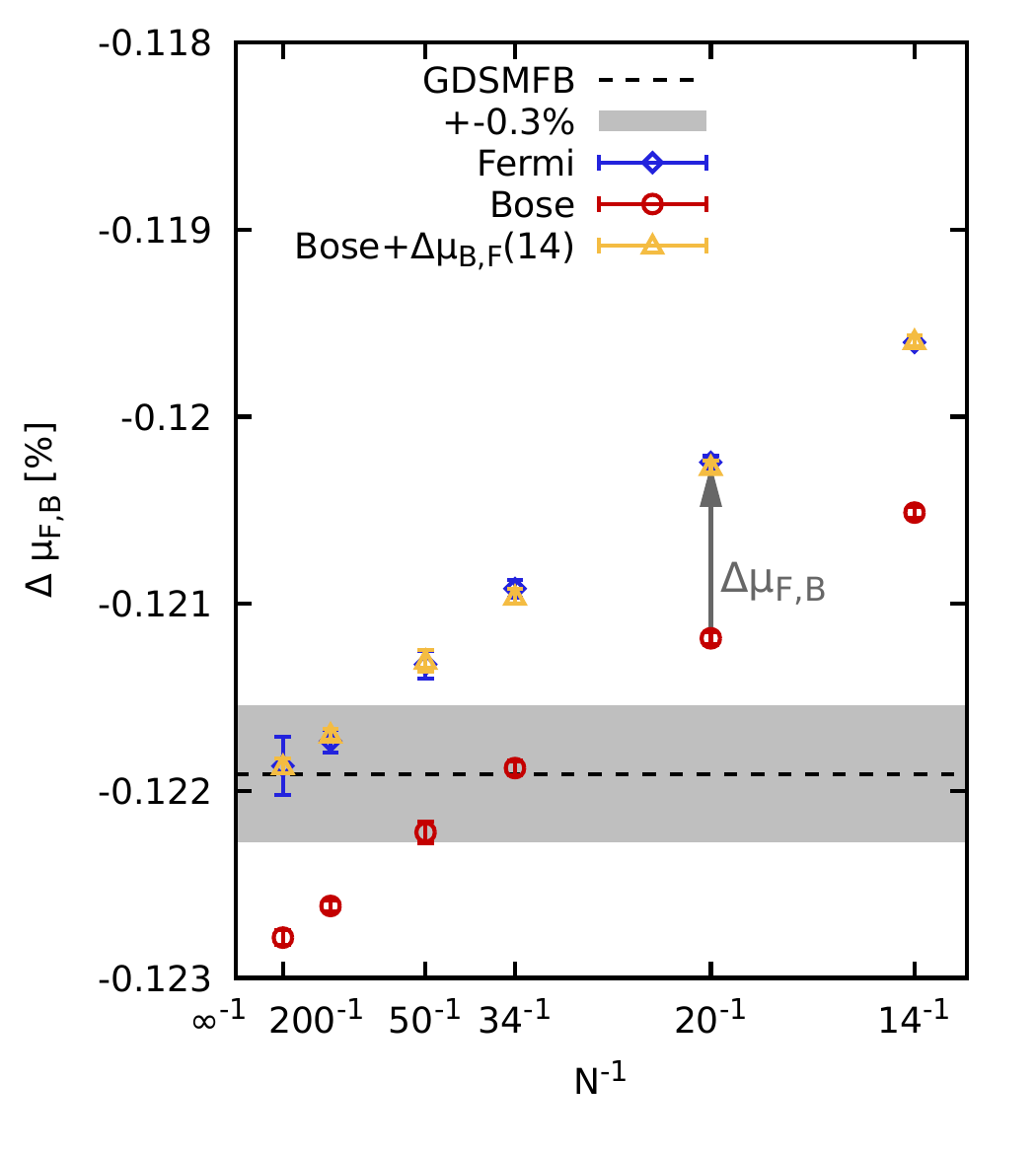}
\caption{\label{fig:Mu_rs10_theta2} Left panel: The dependence of the chemical potential, for the unpolarized UEG with $r_s=10$ and $\Theta=2$, on the number of simulated electrons $N$. Dashed black line: result from the parametrization of $f_\textnormal{xc}(r_s,\Theta)$ by Groth \emph{et al.}~\cite{groth_prl} (GDSMFB) via Eq.~(\ref{eq:mu_from_rs_theta}), and subsequently adding $\mu_0$. Green crosses: results obtained from the direct free energy estimator [Eq.~(\ref{eq:mu_final})]. Blue diamonds: results obtained from the mixed-ensemble histogram estimator [Eq.~(\ref{eq:histrogram_estimator})]. Yellow squares: results with a finite-size corrected non-interacting contribution. Dotted yellow lines: linear fits for $N\geq20$ motivated by the theoretical size-convergence behavior of $\mu_\textnormal{xc}(N,V,\beta)$, see Refs.~\cite{Siepmann_1992,Herdman_PRB_2014} and Eq.~(\ref{eq:mu_TDL}) below. Right panel: The quantum statistics contribution $\Delta\mu_{\textnormal{F},\textnormal{B}}$ is nearly independent of $N$ at the present conditions, facilitating high-fidelity results (yellow triangles) for up to $N\sim200$ electrons.}
\end{figure*} 

All PIMC results that are presented in this work have been computed using the open-source \texttt{ISHTAR} code~\cite{ISHTAR}, and were obtained for a fully unpolarized system. We use $M=50-200$ imaginary-time propagators within the \emph{primitive factorization} $e^{-\epsilon\hat{H}}\approx e^{-\epsilon\hat{K}}e^{-\epsilon\hat{W}}$, and the convergence with the number of time slices $M=\beta/\epsilon$ has been carefully checked. More sophisticated factorization schemes~\cite{sakkos_JCP_2009} and their application to Fermi systems~\cite{Chin_PRE_2015,Dornheim_NJP_2015} have been presented in the literature, but they are not required here. Note that we do not impose any nodal restrictions~\cite{Ceperley1991}, which makes our simulations very costly due to the fermion sign problem~\cite{dornheim_sign_problem}, but exact within the given Monte Carlo error bars.

\subsection{Verification and comparison of estimators\label{sec:verification}}

In the left panel of Fig.~\ref{fig:Mu_rs10_theta2}, we show PIMC simulation results for the total canonical chemical potential $\mu(N,V,\beta)$ as a function of the total number of electrons for the unpolarized UEG at $r_s=10$ and $\Theta=2$. The dashed black line has been obtained from the parametrization of $f_\textnormal{xc}(r_s,\Theta)$ by Groth \emph{et al.}~\cite{groth_prl} (GDSMFB), and has been included as a reference; the shaded grey area shows an interval of $\pm0.3\%$ around this result and serves as a guide to the eye. The green crosses have been obtained by directly evaluating the differences in the free energy [Eqs.~(\ref{eq:Mu_from_F}) and (\ref{eq:mu_final})] and appear to converge towards the GDSMFB reference result. The blue diamonds show independent PIMC results from the histogram estimator [Eq.~(\ref{eq:histrogram_estimator}), with $\mu_\textnormal{GC}=-0.1$] and are in excellent agreement with the independent $\eta$-ensemble results for all $N$ within the given Monte Carlo error bars. Importantly, the latter are considerably smaller for the histogram estimator, making it the method of choice throughout the remainder of this paper.

The yellow squares have been computed from the blue diamonds by finite-size correcting the non-interacting part, i.e., by adding the corresponding finite-size correction
\begin{eqnarray}\label{eq:delta_mu0}
    \Delta\mu_0(N,V,\beta) = \mu_0^\textnormal{TDL} - \mu_0(N,V,\beta)\ .
\end{eqnarray}
Interestingly, adding the correction Eq.~(\ref{eq:delta_mu0}) to the raw PIMC results does not diminish the magnitude of the dependence on the system size, but changes its sign. Nevertheless, adding $\Delta\mu_0(N,V,\beta)$ is very advantageous in practice as the asymptotic dependence of the XC-contribution to $\mu$ on the number of electrons $N$ has been presented in the literature~\cite{Siepmann_1992,Herdman_PRB_2014},
\begin{eqnarray}\label{eq:mu_TDL}
    \Delta\mu_\textnormal{xc}(N,V,\beta) &=& \frac{1}{2N}\left(\frac{\partial P}{\partial n} \right)
    \left\{
1 - \frac{1}{\beta}\left[
\left(\frac{\partial n}{\partial P}\right)\right.\right. \\ & & \left.\left. +n\left(\frac{\partial^2 P}{\partial n^2}\right)\left(\frac{\partial n}{\partial P}\right)^{2}
\right]\right\}+\mathcal{O}(N^{-2})\nonumber\ ,
\end{eqnarray}
with $P$ being the total pressure. The dotted yellow line shows a corresponding linear fit for $N\geq20$, which nicely captures the functional dependence of the PIMC results. Eq.~(\ref{eq:mu_TDL}) thus constitutes a reliable route to extrapolate our PIMC results for $\mu_\textnormal{xc}(N,V,\beta)$ to the appropriate thermodynamic limit without the need for any external finite-size correction.

The right panel of Fig.~\ref{fig:Mu_rs10_theta2} is devoted to the analysis of quantum statistics effects. Specifically, the blue diamonds and red circles show our raw PIMC results for Fermi-Dirac and Bose-Einstein statistics, respectively. In addition, the yellow triangles have been obtained by adding to the red circles a constant quantum statistics correction $\Delta\mu_\textnormal{B,F}$ that has been computed for $N=14$ electrons. Evidently, the thus obtained data points are in perfect agreement with the full fermionic PIMC results, but with a substantially reduced statistical error bar. Naturally, the impact of quantum statistics becomes more pronounced for higher density and lower temperatures, and the utility of a corresponding decomposition [cf.~Eq.~(\ref{eq:mu_final})] is investigated in the following.

\subsection{Decomposition and extrapolation to the thermodynamic limit\label{sec:extrapolation}}

\begin{figure*}\centering
\includegraphics[width=0.32\textwidth]{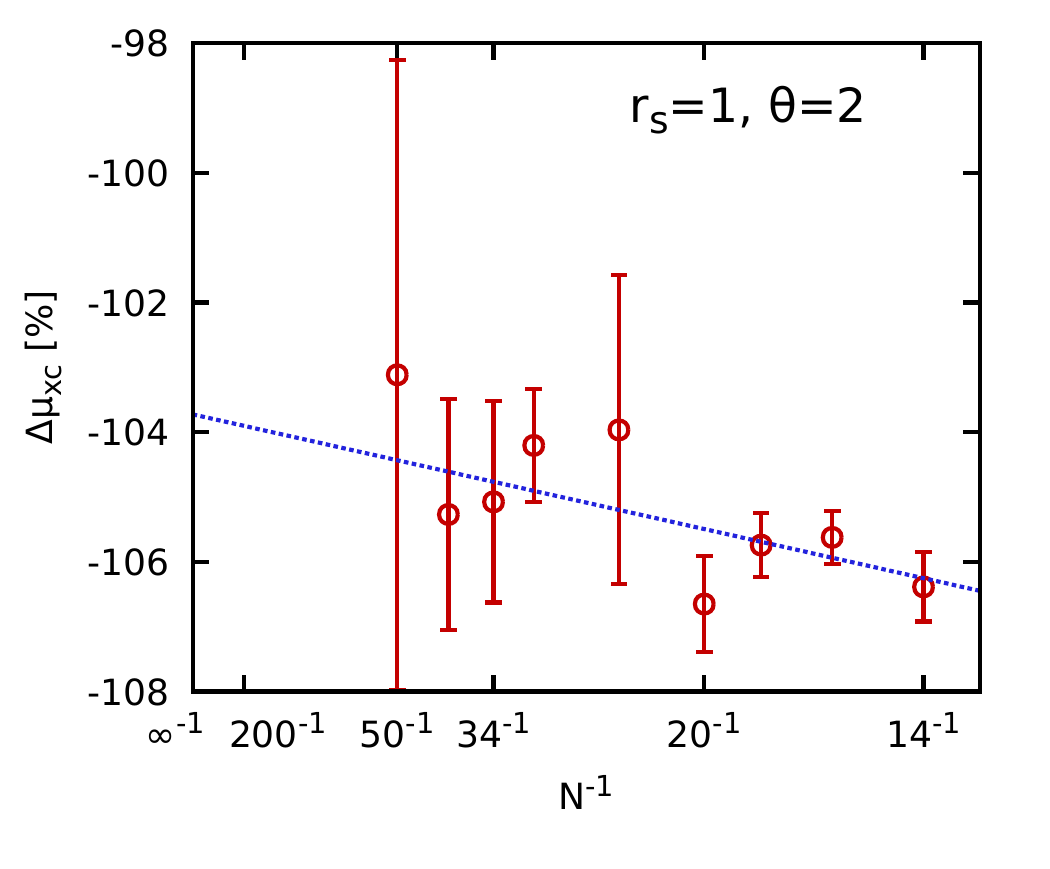}\includegraphics[width=0.32\textwidth]{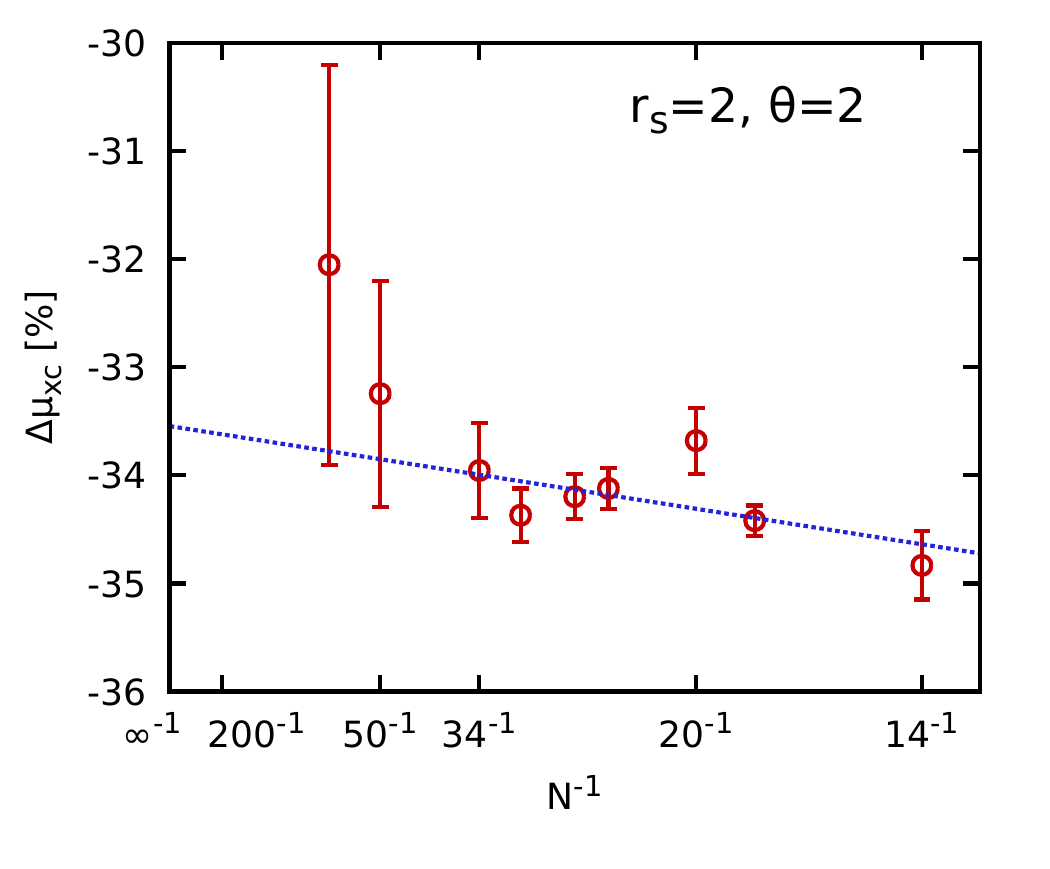}\includegraphics[width=0.32\textwidth]{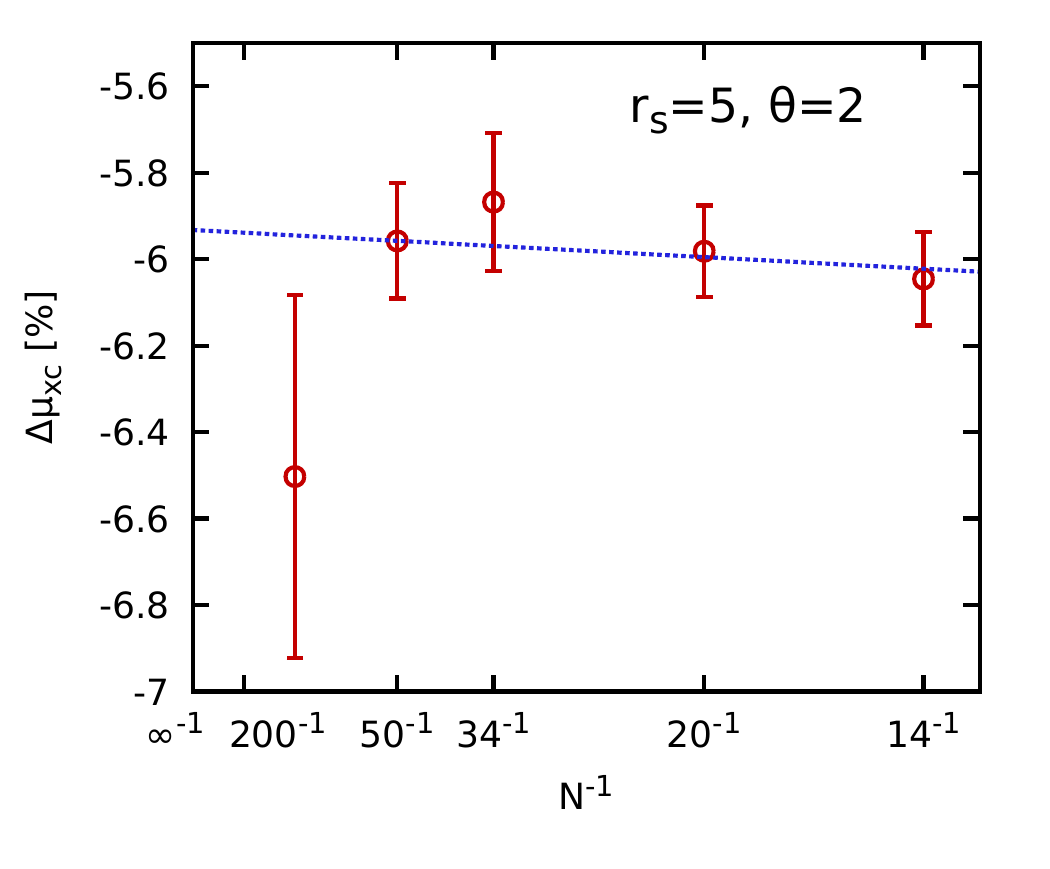}\\\includegraphics[width=0.32\textwidth]{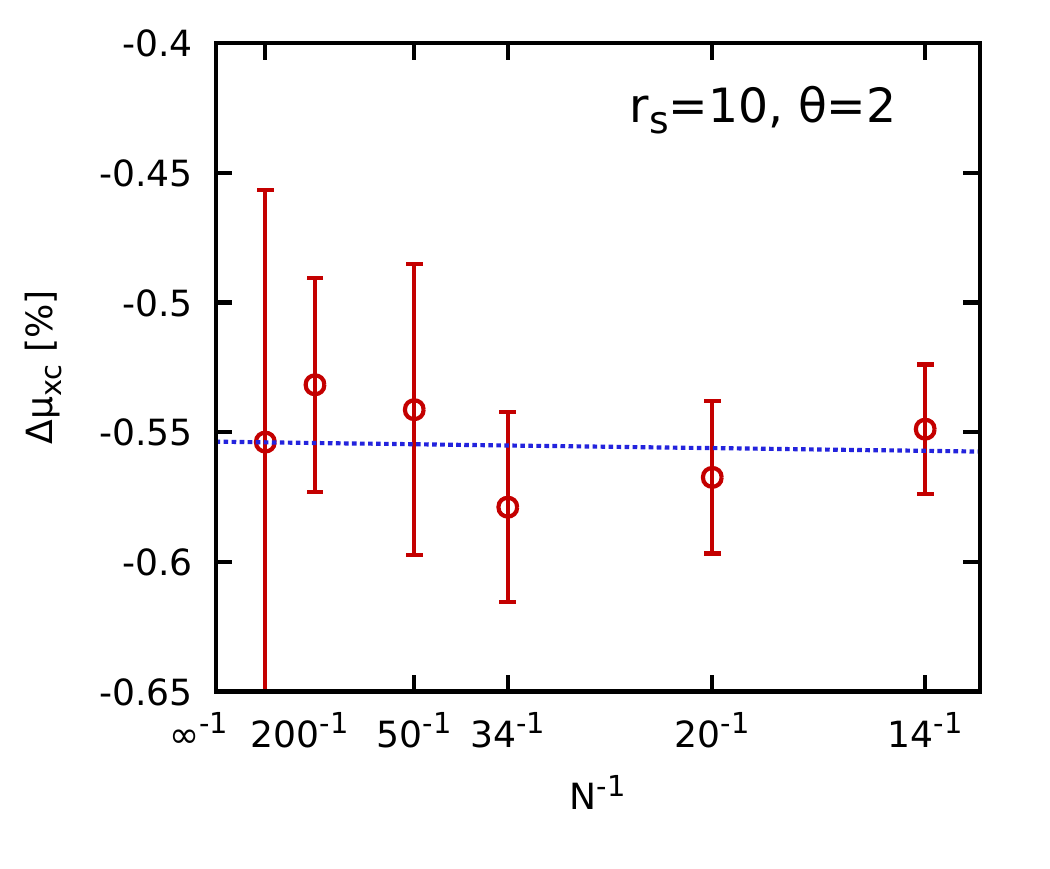}\includegraphics[width=0.32\textwidth]{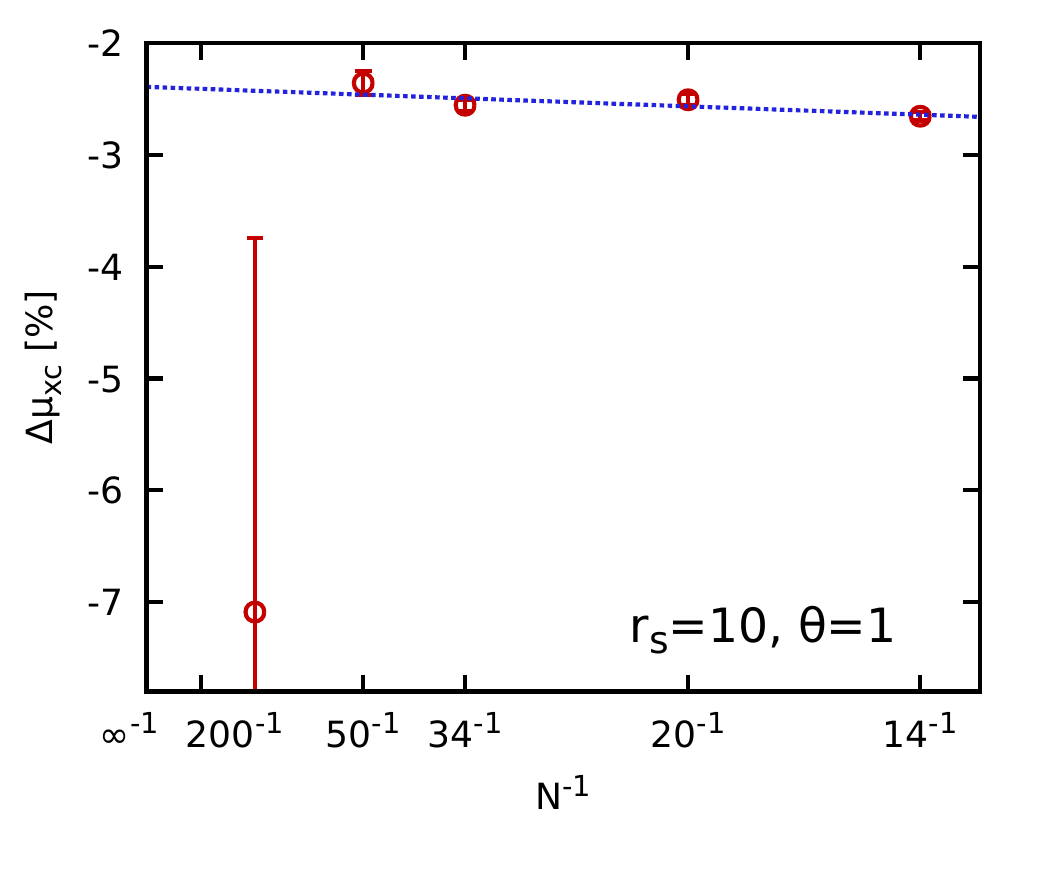}\includegraphics[width=0.32\textwidth]{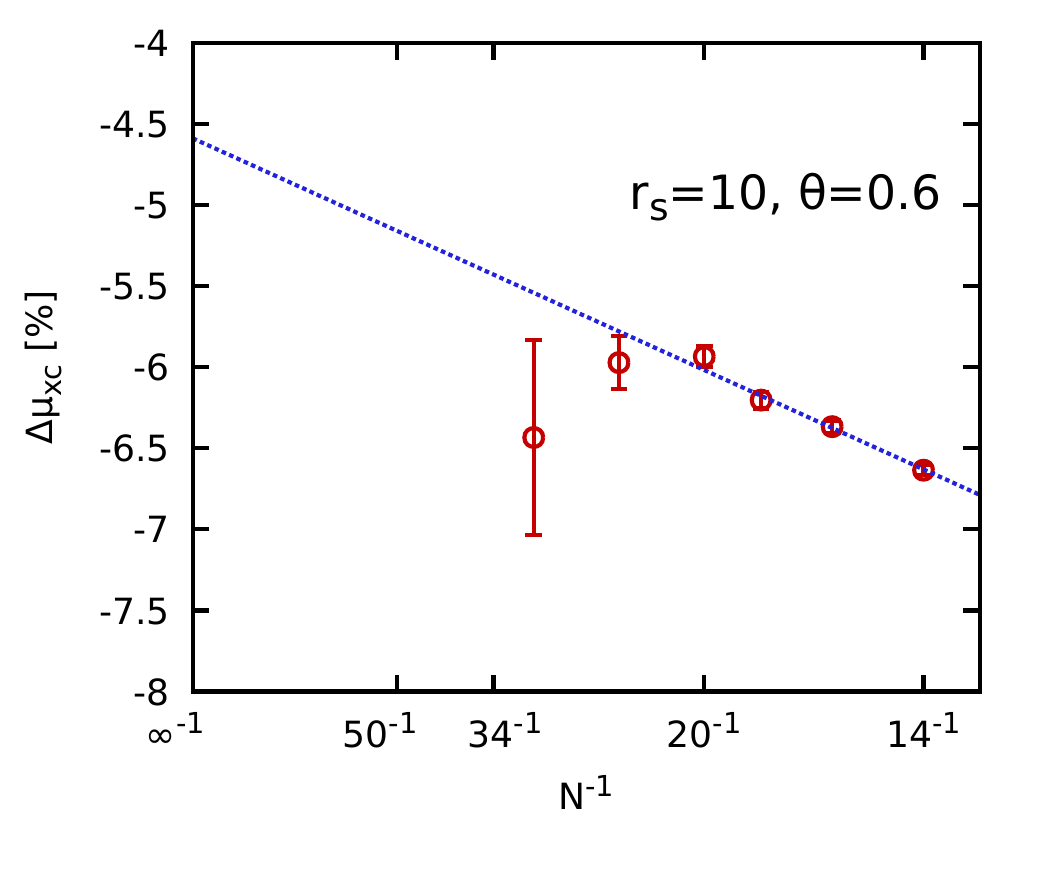}
\caption{\label{fig:delta_statistics} \emph{Ab initio} PIMC results for the relative quantum statistical contribution $\Delta\mu_\textnormal{B,F}$ (that is computed with respect to the GDSMFB~\cite{groth_prl} results for $\mu_\textnormal{xc}$) as a function of system size $N$ for different $r_s$--$\Theta$ combinations. Red circles: raw PIMC results; dotted blue lines: empirical linear fits.
}
\end{figure*} 

In Fig.~\ref{fig:delta_statistics}, we show the quantum statistical contribution $\Delta\mu_\textnormal{F,B}$, i.e., the difference between the chemical potential obtained for Fermi-Dirac and Bose-Einstein statistics, as a function of system size for various combinations of the density and temperature. Note that we normalize these differences by the $\mu_\textnormal{xc}$ computed from the GDSMFB parametrization~\cite{groth_prl} for convenience.
The top row has been obtained for $\Theta=2$ and $r_s=1$ (left), $r_s=2$ (center), and $r_s=5$ (right), i.e., for typical WDM parameters. Remarkably, we find differences exceeding $100\%$ in $\mu_\textnormal{xc}$ between fermions and bosons for $r_s=1$, which clearly emphasizes the importance of quantum degeneracy effects for the description of WDM. At the same time, the dependence of $\Delta\mu_\textnormal{F,B}$ on $N$ is relatively weak for all depicted cases. This also holds for the bottom row, where we show results for $r_s=10$, which corresponds to the boundary of the strongly coupled electron liquid regime~\cite{dornheim_electron_liquid,dornheim_dynamic}, for $\Theta=2$ (left), $\Theta=1$ (center) and $\Theta=0.6$ (right).

This observation is of great practical value, since PIMC simulations of large bosonic systems are comparatively unproblematic, as they are not subject to the exponential increase in compute time due to the fermion sign problem.
We thus propose to use bosonic PIMC results for $\mu_\textnormal{B}(N,V,\beta)$ as a basis, and to correct them with PIMC results for $\mu_\textnormal{F,B}$ that can be estimated reliably for a significantly smaller number of particles. In practice, we use an empirical linear model for $\Delta\mu_\textnormal{F,B}(N,V,\beta)$, which is included as the dotted blue lines in Fig.~\ref{fig:delta_statistics}; we find excellent agreement between the PIMC results and these fits for all investigated cases, and over the entire investigated range of system sizes $N$.

\begin{figure*}\centering
\includegraphics[width=0.32\textwidth]{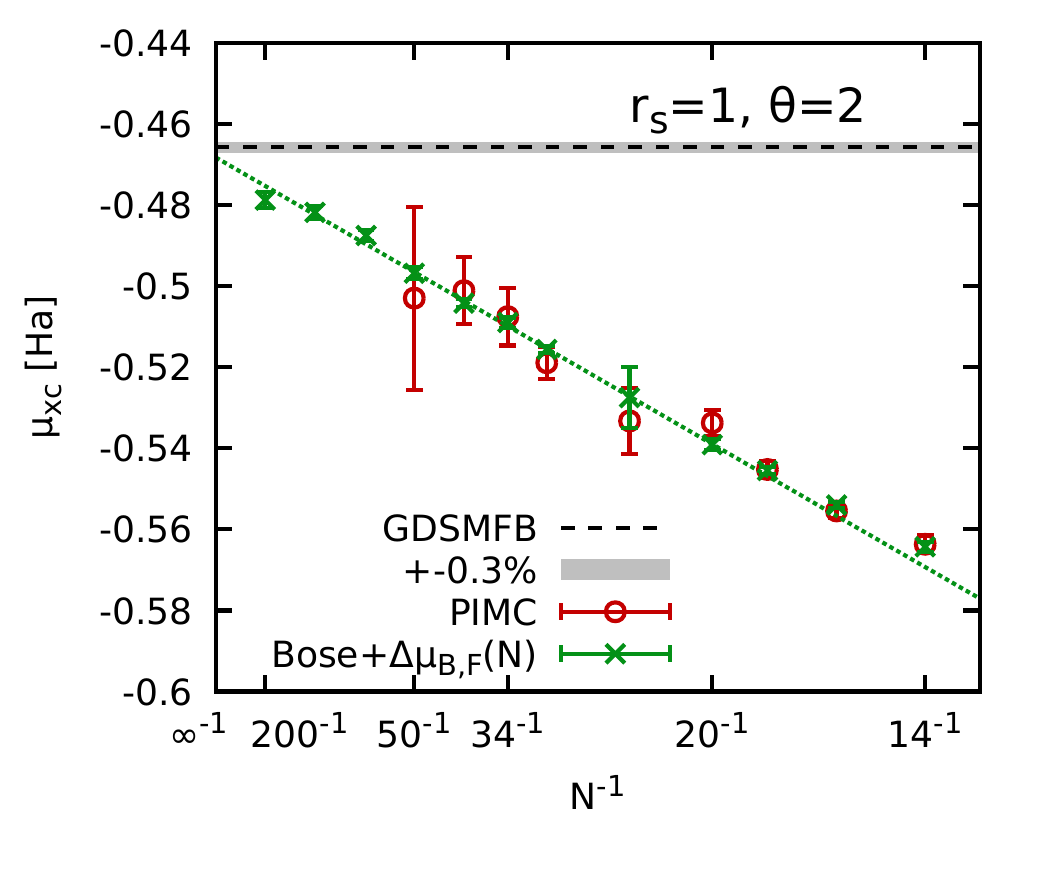}\includegraphics[width=0.32\textwidth]{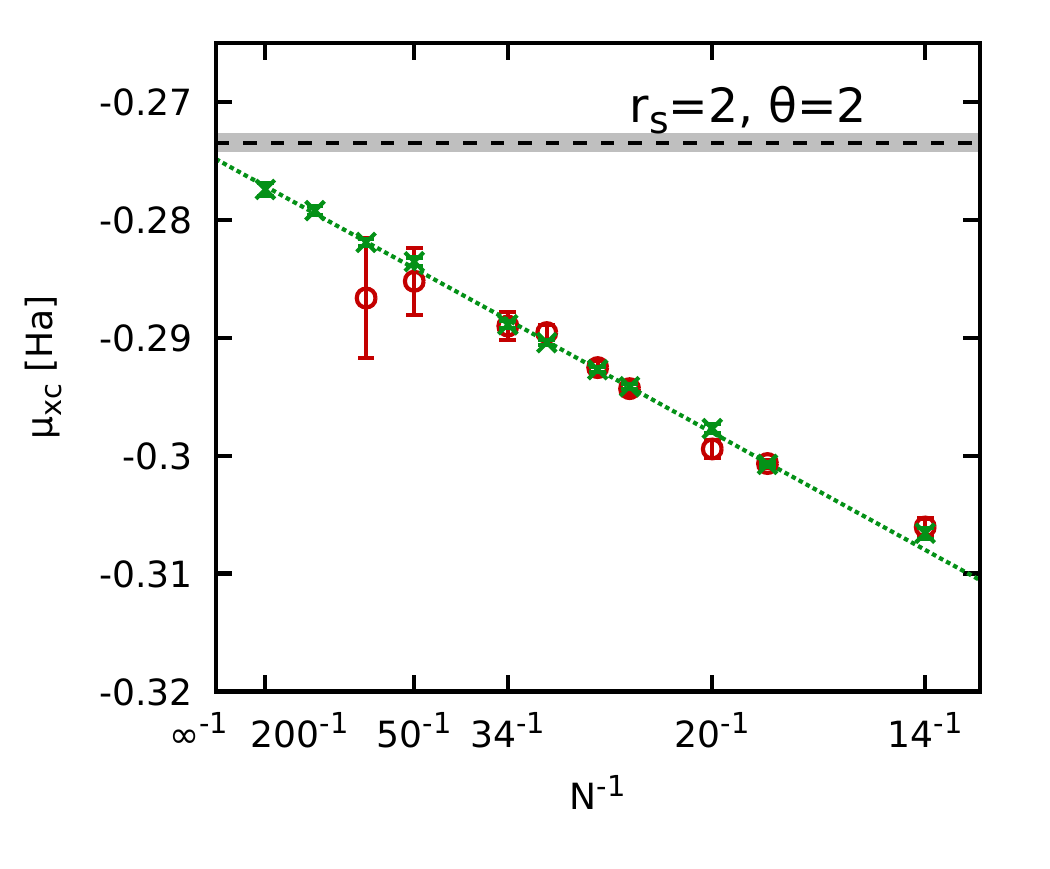}\includegraphics[width=0.32\textwidth]{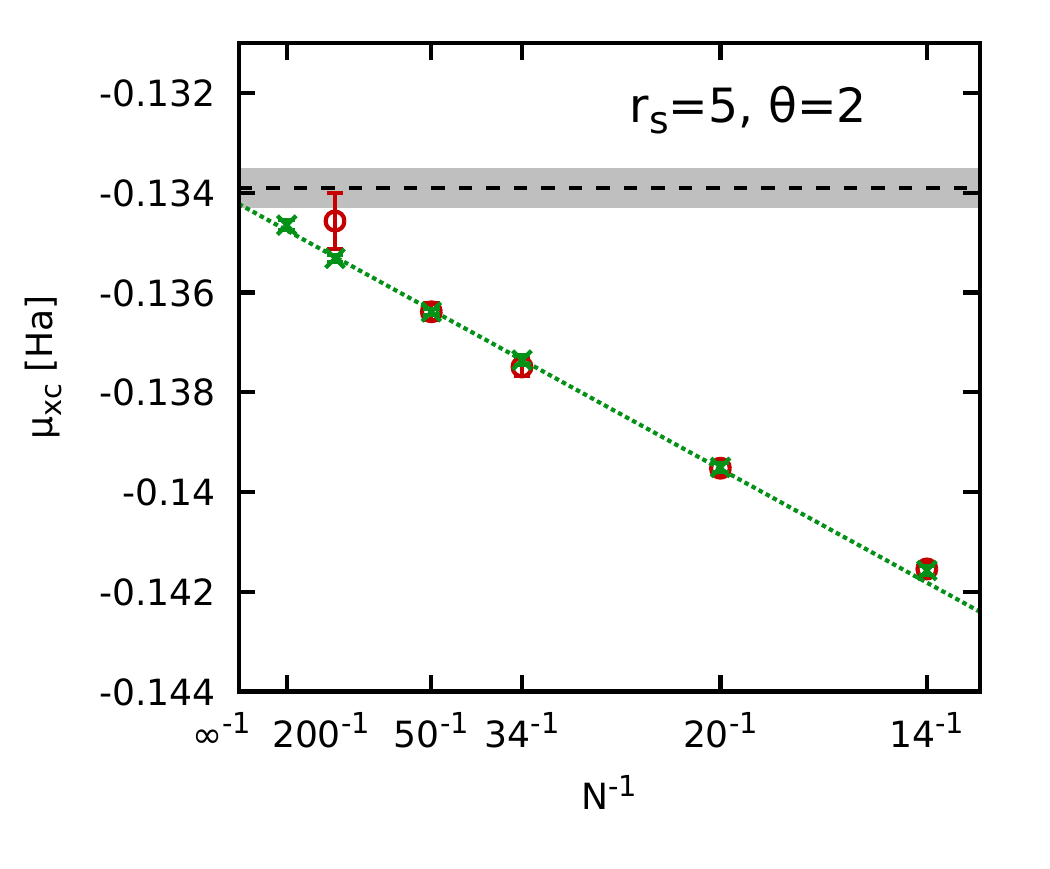}\\\includegraphics[width=0.32\textwidth]{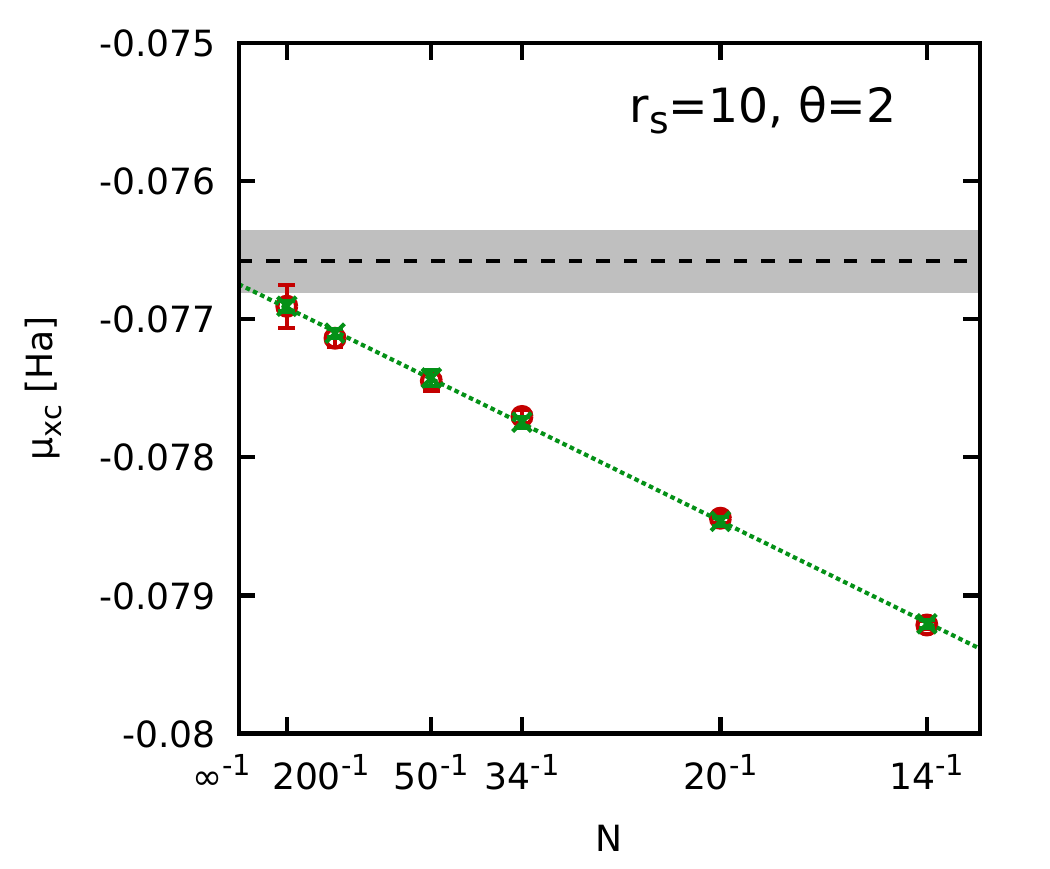}\includegraphics[width=0.32\textwidth]{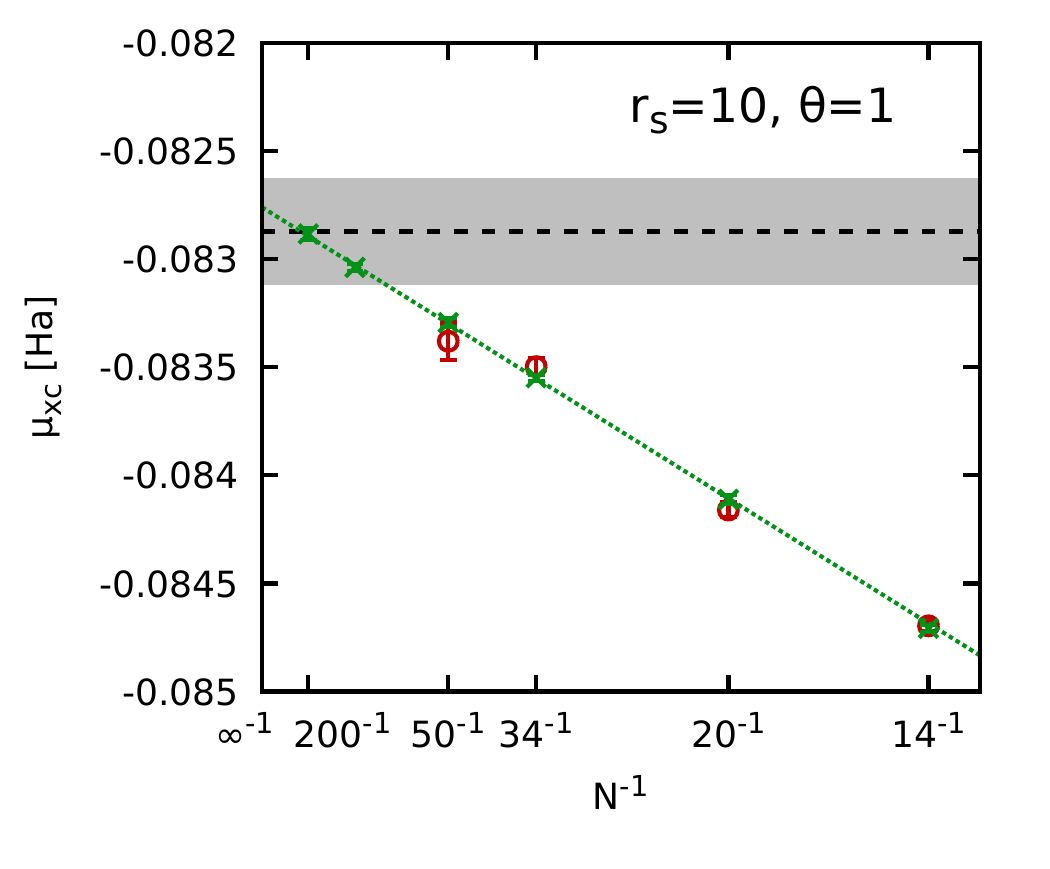}\includegraphics[width=0.32\textwidth]{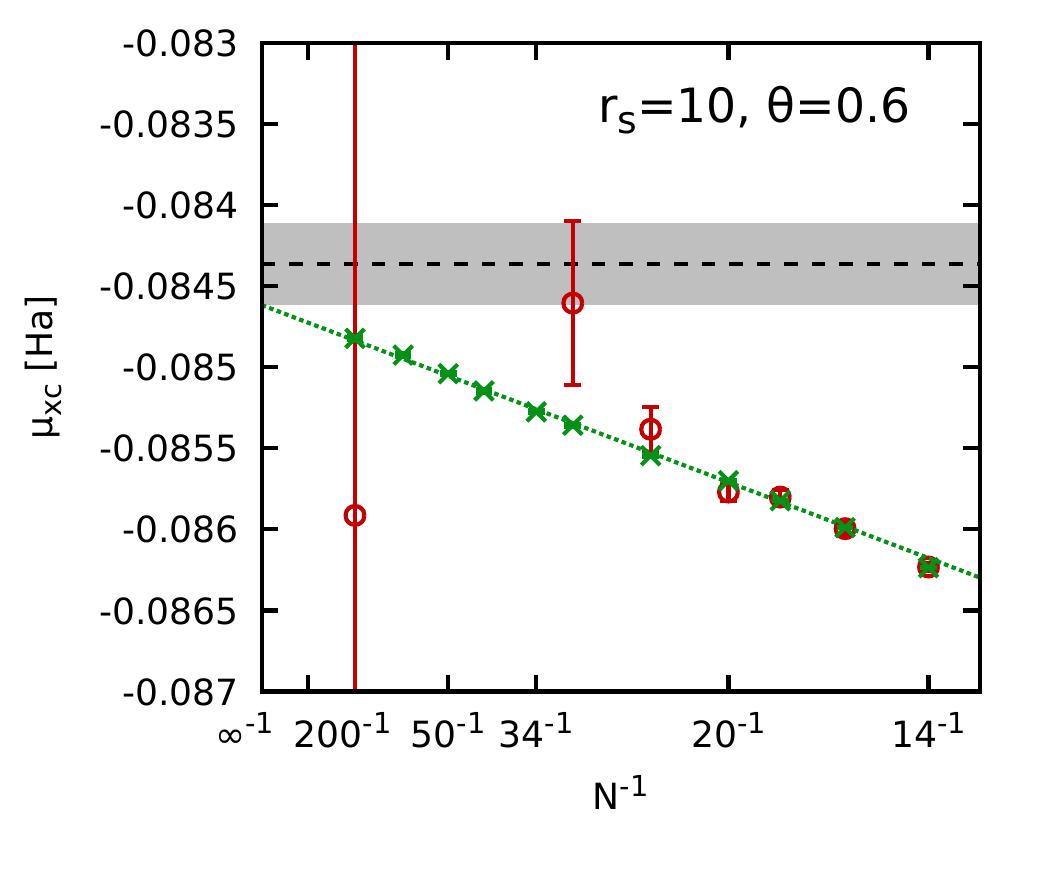}
\caption{\label{fig:mu_xc_extrapolation} Extrapolation of the \emph{ab initio} PIMC results for the XC-contribution to the chemical potential $\mu_\textnormal{xc}$ to the TDL as a function of system size for a variety of $r_s$--$\Theta$ combinations. Red circles: raw fermionic PIMC results; green crosses: bosonic PIMC results corrected with PIMC estimates for the quantum statistics correction $\Delta\mu_\textnormal{B,F}(N)$ as it has been depicted in Fig.~\ref{fig:delta_statistics}; dotted green lines: linear fits for $N\geq20$, cf.~Eq.~(\ref{eq:mu_TDL}); dashed gray line: $\mu_\textnormal{xc}(r_s,\Theta)$ computed from the parametrization of $f_\textnormal{xc}(r_s,\Theta)$ by Groth \emph{et al.}~\cite{groth_prl} (GDSMFB), cf.~Eq.~(\ref{eq:mu_from_rs_theta}); shaded gray area: nominal uncertainty interval of GDSMFB of $\pm0.3\%$.
}
\end{figure*} 

Let us next consider Fig.~\ref{fig:mu_xc_extrapolation}, where we investigate the extrapolation of $\mu_\textnormal{xc}$ to the TDL for the same parameters as in Fig.~\ref{fig:delta_statistics}. Overall, we find the most pronounced dependence on the system size for the highest density ($r_s=1$ and $\Theta=2$), where the finite-size effect for $N=14$ attains roughly one half of $\mu_\textnormal{xc}$ in the TDL. In contrast, finite-size effects decrease towards lower temperatures and larger $r_s$, which is consistent with the investigation of the finite-size effects of other observables in the available literature~\cite{dornheim_prl,review,Dornheim_JCP_2020,Dornheim_F_Follow_up}.
More specifically, the red circles show our raw fermionic PIMC results; these are very accurate for small $N$, but exhibit substantially increasing statistical error bars with increasing $N$ due to the fermion sign problem, in particular at high densities and low temperatures. In stark contrast, the green crosses that have been obtained by adding to the bosonic PIMC results the quantum statistical correction as it has been explained in the discussion of Fig.~\ref{fig:delta_statistics} above, are not afflicted by this exponential scaling and remain accurate even for $N=200$. This allows us to reliably extrapolate to the TDL, see the dotted green curves that correspond to linear fits for $N\geq20$  
motivated by Eq.~(\ref{eq:mu_TDL}). We again stress that this procedure does not involve any of the usual external finite-size correction procedures that have been applied for the construction of both GDSMFB and corrKSDT (as well as to the original KSDT~\cite{ksdt}, albeit in a truncated first-order form~\cite{Brown_PRL_2013,dornheim_prl}).





\begin{figure*}\centering
\includegraphics[width=0.32\textwidth]{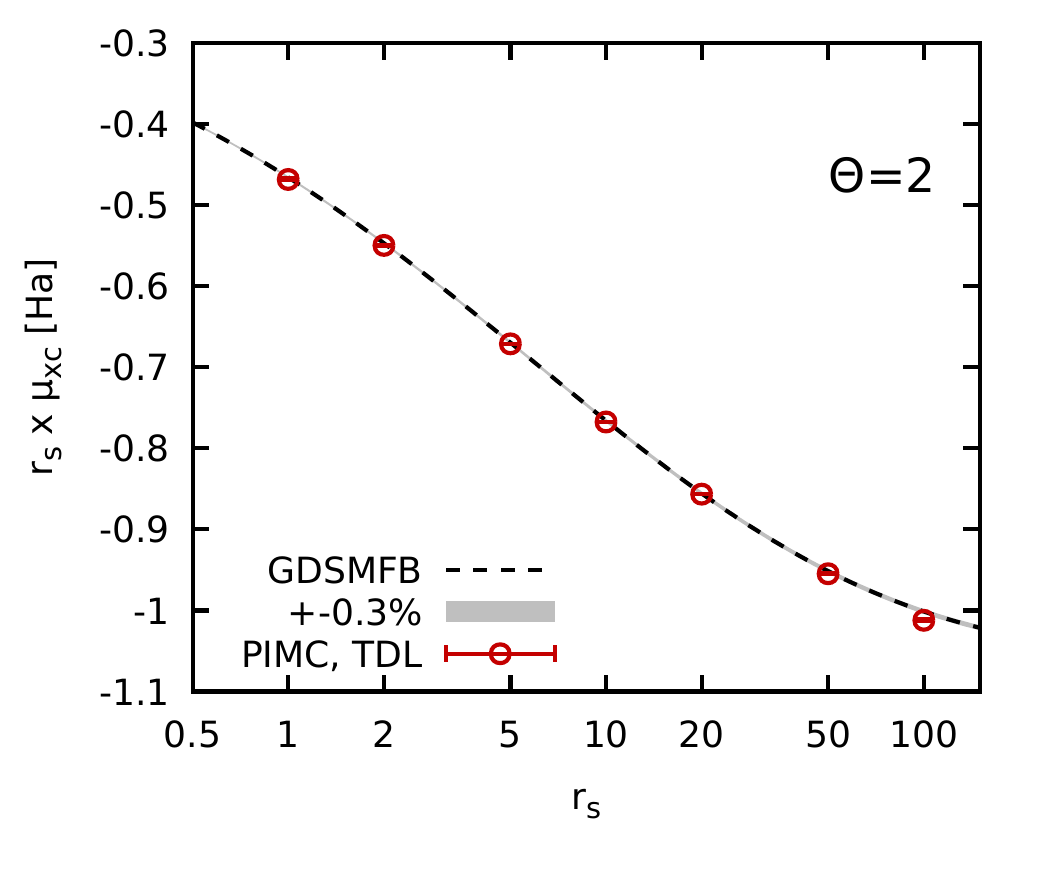}\includegraphics[width=0.32\textwidth]{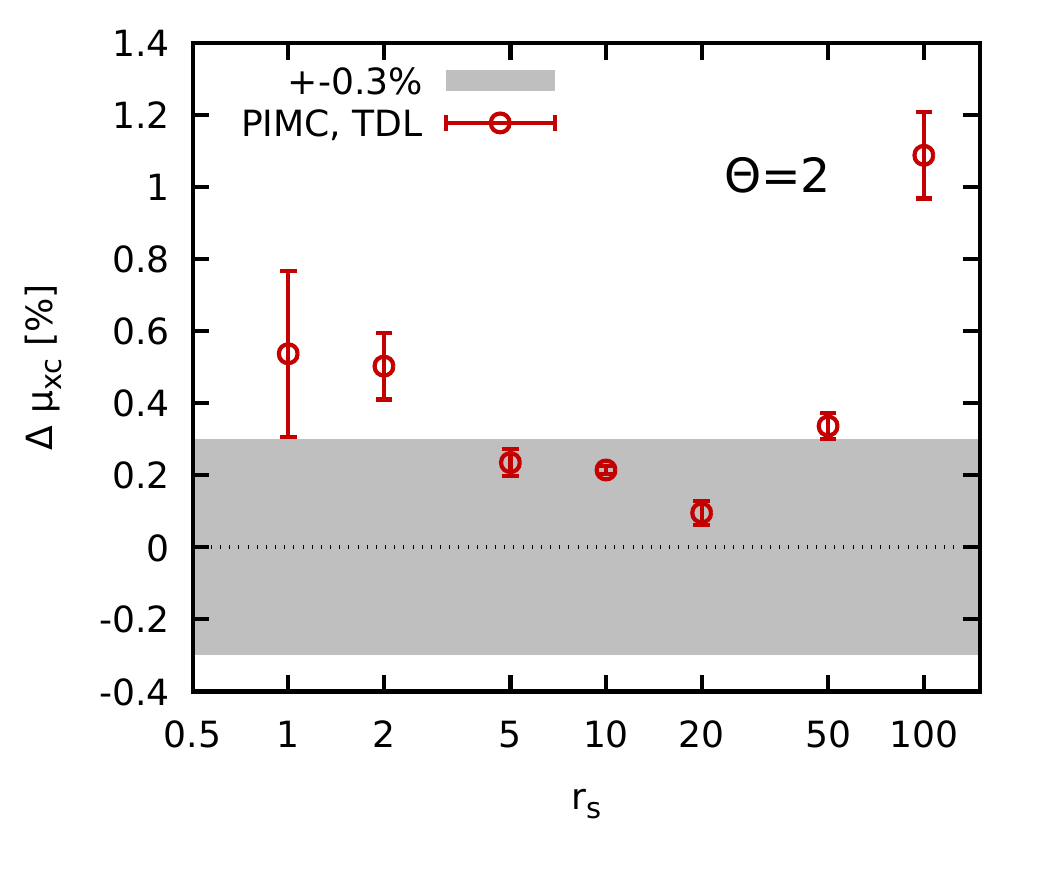}\includegraphics[width=0.32\textwidth]{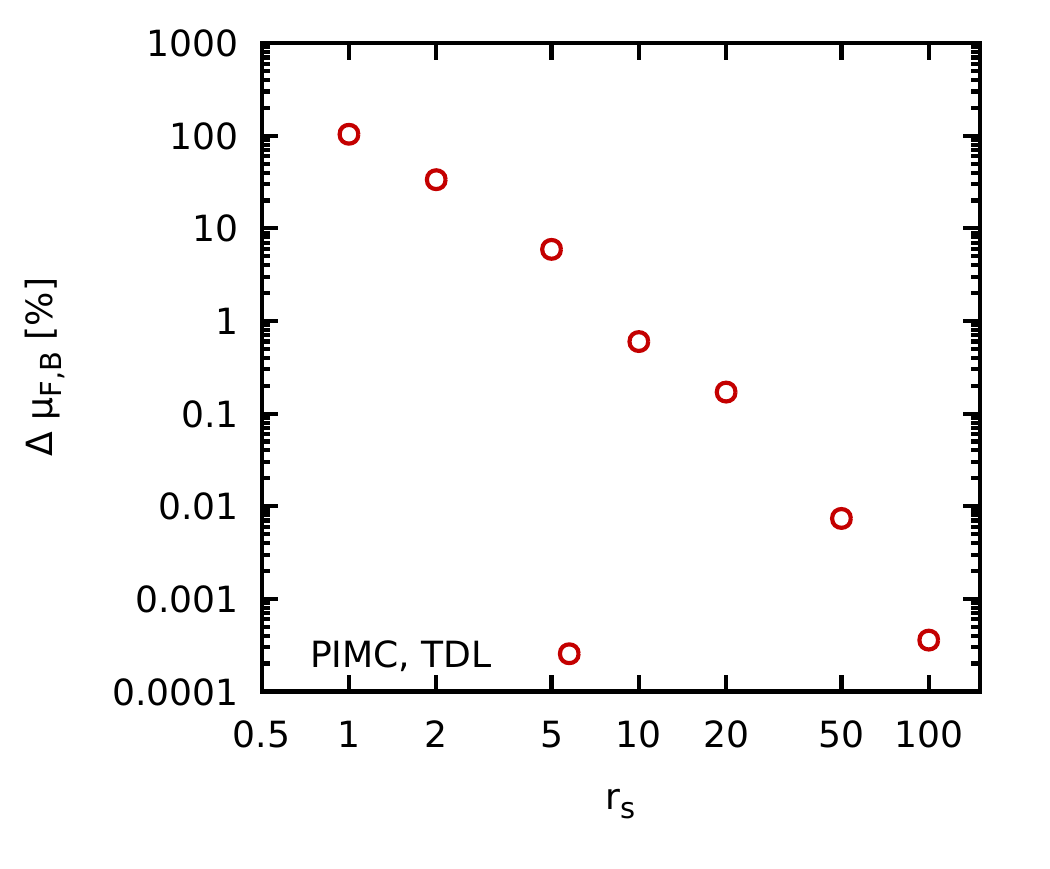}\\\includegraphics[width=0.32\textwidth]{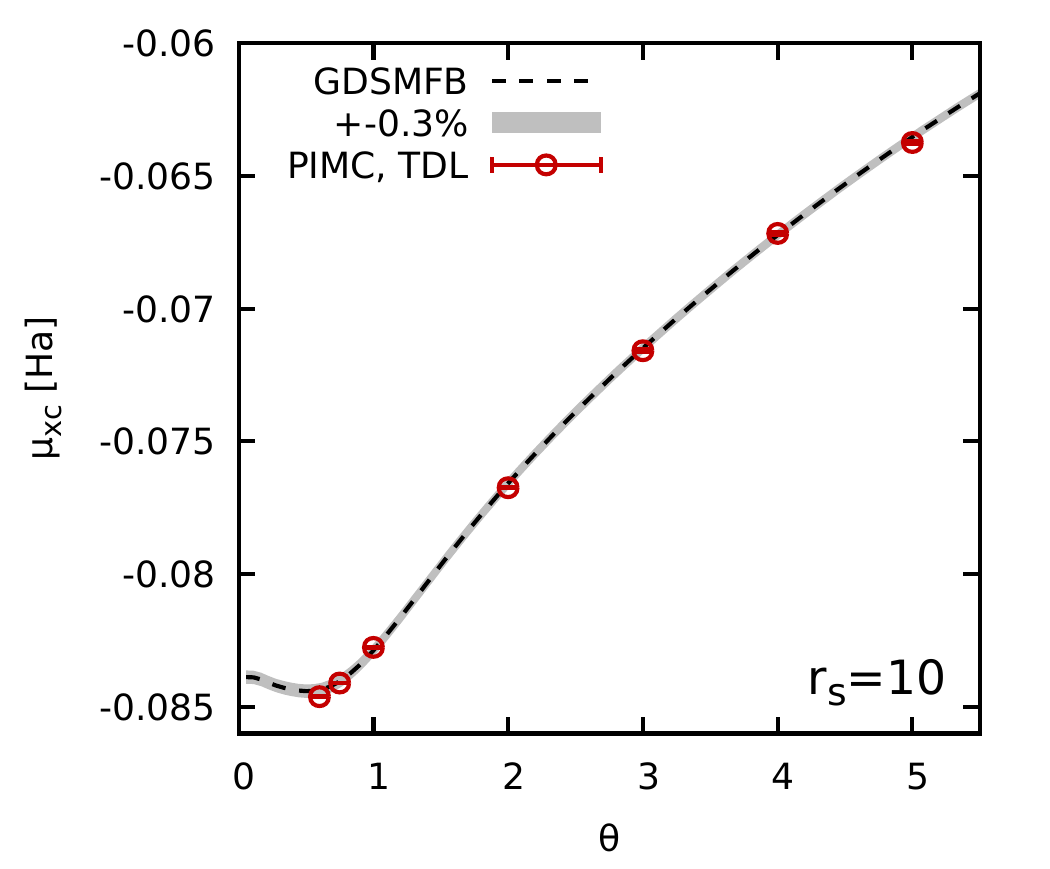}\includegraphics[width=0.32\textwidth]{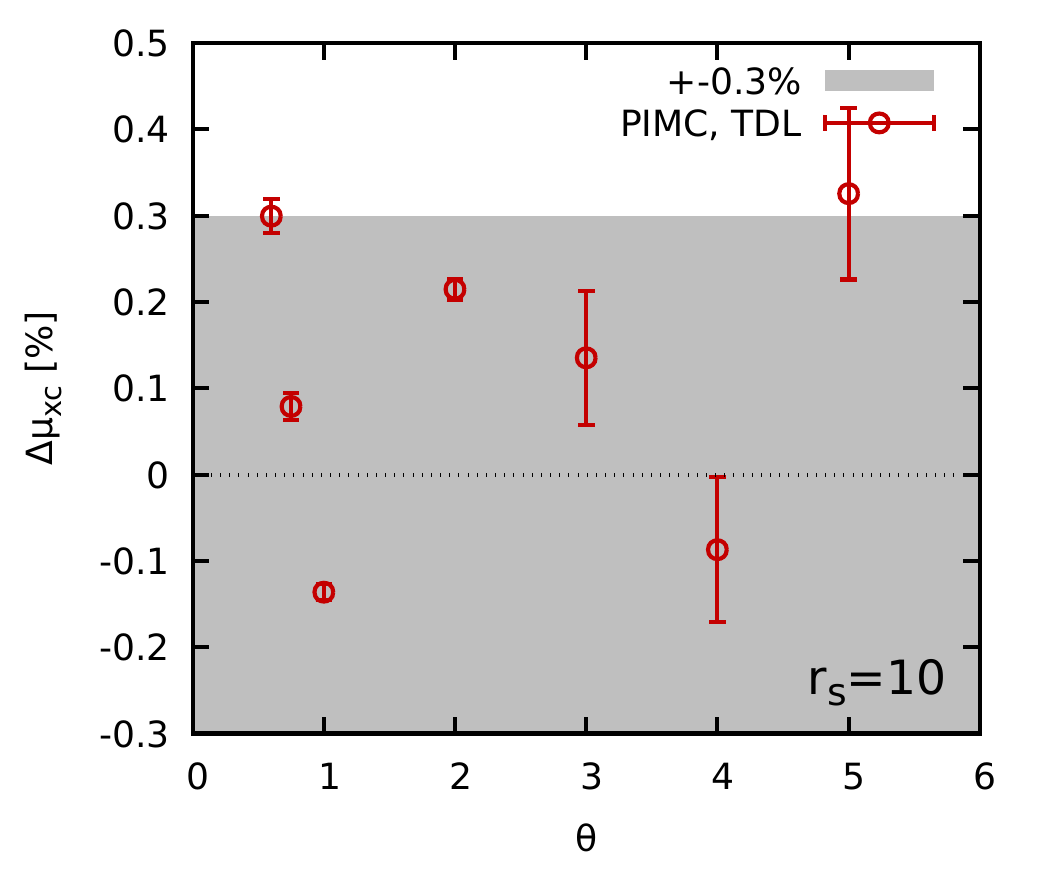}\includegraphics[width=0.32\textwidth]{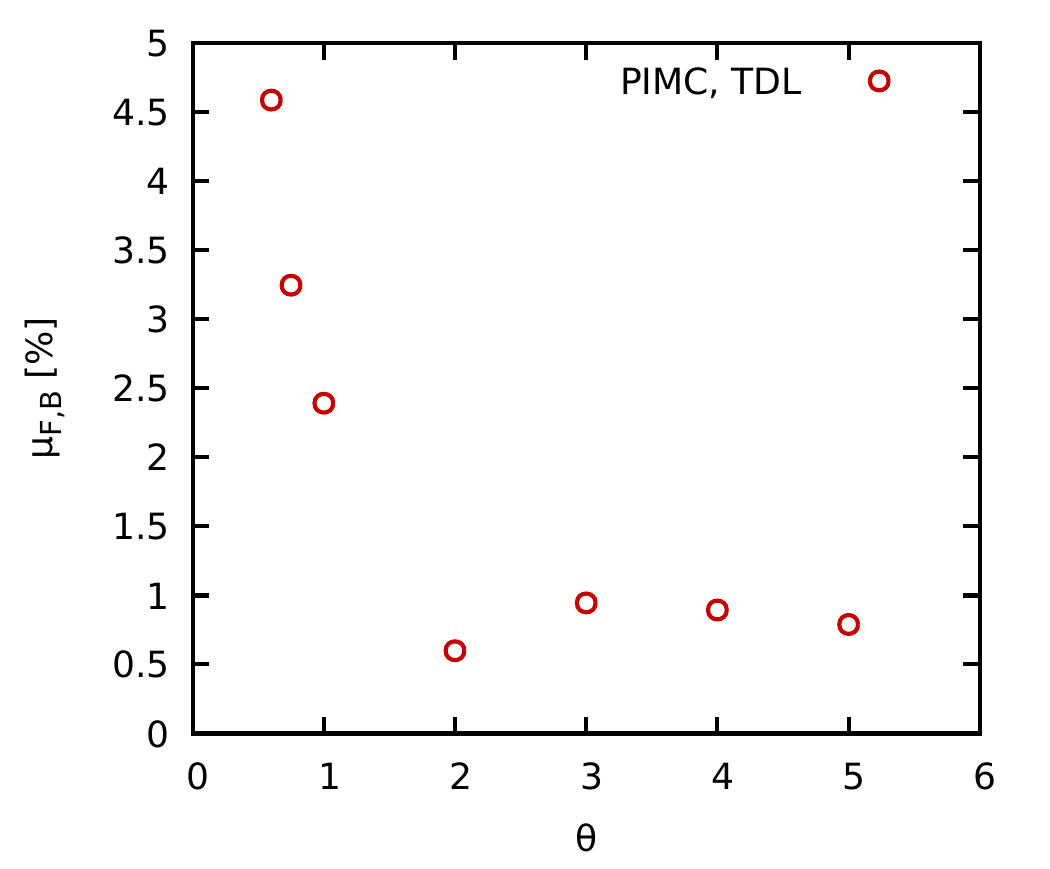}
\caption{\label{fig:mu_xc_rs10_theta} Top (bottom): $r_s$-dependence ($\Theta$-dependence) of the XC-contribution to the chemical potential $\mu_\textnormal{xc}$. Left: comparing the present extrapolation to the TDL (red circles) to $\mu_\textnormal{xc}$ computed from the parametrization of $f_\textnormal{xc}(r_s,\Theta)$ by Groth \emph{et al.}~\cite{groth_prl} (GDSMFB, dashed black) via Eq.~(\ref{eq:mu_from_rs_theta}). Center: relative deviation between the present work and GDSMFB, with the shaded black area being a nominal uncertainty interval of $\pm0.3\%$. Right: relative impact of the quantum statistical contribution $\mu_\textnormal{B,F}$ in relation to $\mu_\textnormal{xc}$.
}
\end{figure*}

\subsection{Comparison to previous parametrization\label{sec:parametrization}}

Let us conclude our investigation by comparing our new extrapolated results for $\mu_\textnormal{xc}$ [see also Table~\ref{tab:UEG}] to the state-of-the-art parmetrization of $f_\textnormal{xc}(r_s,\Theta)$ by Groth \emph{et al.}~\cite{groth_prl} (GDSMFB). We note that the corrected KSDT functional (corrKSDT) is largely indistinguishable from GDSMFB~\cite{status}, so that our conclusions can be expected to hold for both parametrizations.

In the top row of Fig.~\ref{fig:mu_xc_rs10_theta}, we first investigate the $r_s$-dependence of $\mu_\textnormal{xc}(r_s,\Theta)$ for a constant value of the degeneracy temperature of $\Theta=2$. The top left panel shows the results for $\mu_\textnormal{xc}$ itself (multiplied by $r_s$ for convenience), with the dashed black line corresponding to GDSMFB and the red circles to our new results. Clearly, both data sets exhibit the same qualitative behavior over the depicted range of densities. To get a more quantitative comparison, we show the relative deviation in the central panel. The largest deviation between GDSMFB and the present work is given by $\Delta\mu_\textnormal{xc}\sim1\%$ at $r_s=100$. Indeed, GDSMFB has been based on configuration PIMC and permutation blocking PIMC input data for $r_s\leq20$, which makes the good agreement so far out of its range of application all the more impressive. A second trend is given by the systematic sign of the observed deviation, which is always positive; in other words, the new PIMC results are always somewhat smaller than the result computed from GDSMFB at $\Theta=2$, even though this deviation is small in magnitude. In fact, GDSMFB boasts a relative accuracy of $\Delta f_\textnormal{xc}/f_\textnormal{xc}\sim0.3\%$~\cite{groth_prl} (which has inspired the shaded gray area in Fig.~\ref{fig:mu_xc_rs10_theta}), but this can be expected to translate into a larger uncertainty for its derivatives. This point has been investigated recently by Karasiev \emph{et al.}~\cite{status}, who found unphysical oscillations in the heat capacity computed from either GDSMFB or corrKSDT.
On the other hand, this discrepancy might also stem from the present work, possibly from the extrapolation to the thermodynamic limit. Indeed, while the asymptotic $N$-dependence seems to empirically hold for $N\geq20$ based on our PIMC data, a small residual second-order effect might induce a systematic bias in $\mu_\textnormal{xc}$, which might also explain the observed small deviations. Here, we cannot conclusively resolve this intriguing question at these conditions; the observed agreement on the order of $\sim0.5\%$ is well within the expected accuracy range for a derivative of GDSMFB and is thus satisfactory. For completeness, we show the relative importance of quantum statistical effects in the right panel. Evidently, it quickly decays with decreasing density where the formation of permutation cycles is suppressed by the increasing degree of Coulomb repulsion, and it becomes negligible in the electron liquid regime.

In the bottom row of Fig.~\ref{fig:mu_xc_rs10_theta}, we investigate the $\Theta$-dependence of $\mu_\textnormal{xc}$ at $r_s=10$, for which GDSMFB exhibits a nontrivial minimum around $\Theta=0.5$. A similar non-monotonic behavior has been observed by Groth \emph{et al.}~\cite{Groth_PRB_2016} for the XC-energy $E_\textnormal{xc}$, where it has been explained as a competition between quantum delocalization and thermal disorder, which increases towards low and high temperatures, respectively. While our PIMC simulations are limited to $\Theta\geq0.6$, we can clearly resolve a lower value for $\mu_\textnormal{xc}$ at $\Theta=0.6$ compared to the zero-temperature limit.  In the center panel of the bottom row, we show again the relative deviation between the present work and GDSMFB. Evidently, all new PIMC data points are in good agreement with the $\pm0.3\%$ interval around the latter, including the lowest temperature point at $\Theta=0.6$.
This regime constitutes the most uncertain part of GDSMFB, where its functional dependence has been determined based on temperature-corrected STLS data points~\cite{groth_prl}.
Finally, the quantum statistical contribution $\mu_\textnormal{B,F}$ that is depicted in the right panel systematically increases towards low temperatures, and attains $\sim4.5\%$ at $\Theta=0.6$ at this relatively low density.

\begin{table*}
\caption{\label{tab:UEG}\emph{Ab initio} PIMC results for the XC-contribution to the chemical potential [in Hartree] extrapolated to the thermodynamic limit. Top half: $r_s$-dependence for $\Theta=2$; bottom half: $\Theta$-dependence for $r_s=10$.}
\begin{ruledtabular}
\begin{tabular}{lllllll}
  $r_s=1$  & $r_s=2$ & $r_s=5$ & $r_s=10$ & $r_s=20$ & $r_s=50$ & $r_s=100$
 \\[+1ex]\colrule\\[-1ex]
$-0.468(1)$ & $-0.2748(2)$ & $-0.13422(5)$ & $-0.0767453(9)$ & $-0.042826(15)$ & $-0.019094(7)$ & $-0.010122(12)$

 \\[+1ex]\colrule\\[-2.35ex]\colrule\\[-1ex]
$\Theta=0.6$ & $\Theta=0.75$ & $\Theta=1$ & $\Theta=2$ & $\Theta=3$ & $\Theta=4$ & $\Theta=5$
 \\[+1ex]\colrule\\[-1ex]
$-0.08462(2)$ & $-0.083989(14)$ & $-0.082759(8)$ & $-0.076745(9)$ & $-0.07159(6)$ & $-0.06717(6)$ & $-0.06374(6)$
\end{tabular}
\end{ruledtabular}
\end{table*}

\section{Summary and Discussion\label{sec:outlook}}

In this work, we have presented extensive new \emph{ab initio} PIMC results for the chemical potential of the warm dense UEG. To this end, we have implemented two independent strategies: i) the direct estimation of $\mu$ as the change in free energy by computing the latter for both $N$ and $N+1$ particles using the recent extended $\eta$-ensemble approach~\cite{dornheim2024directfreeenergycalculation,Dornheim_F_Follow_up} and ii) the estimation of $\mu$ from the ratio of the corresponding partition functions based on PIMC simulations with a varying particle number. In practice, both routes give the same results, with the histogram estimator being the more efficient choice. A great advantage of the chemical potential compared to other thermodynamic observables such as the free energy $F$, internal energy $E$, or kinetic/interaction energy $K$ and $W$ is its a-priori known asymptotic dependence on the system size $N$~\cite{Siepmann_1992,Herdman_PRB_2014}. In combination with the decomposition of $\mu$ into an easy to compute bosonic contribution $\mu_\textnormal{B}$ and an additional quantum statistical correction $\mu_\textnormal{B,F}$ that can be estimated from relatively small system sizes, this has facilitated the reliable extrapolation to the thermodynamic limit based on PIMC results for up to $N=200$ electrons. This has been achieved without the need for any additional external finite-size corrections, as those employed for the construction of previous PIMC based parametrizations of the UEG EOS.

Overall, we find a very good agreement between the present results and the XC-contribution to $\mu$ computed from the XC-free energy parametrization by Groth \emph{et al.}~\cite{groth_prl}. More specifically, we have found a maximum deviation of $\Delta\mu_\textnormal{xc}\sim1\%$ at $r_s=100$, which, however, is well outside the nominal applicability range of GDSMFB. For $r_s\leq20$, we observe a maximum deviation of $\Delta\mu_\textnormal{xc}\sim0.5\%$ at $r_s=1,2$, with a small but systematic trend for all densities at $\Theta=2$. This could fully be explained by the nominal confidence interval of $\Delta f_\textnormal{xc}\sim0.3\%$ reported by Groth \emph{et al.}~\cite{groth_prl}, which can easily be exceeded by the evaluation of the derivatives both with respect to $\Theta$ and $r_s$ that are required for the estimation of $\mu_\textnormal{xc}$ [cf.~Eq.~(\ref{eq:mu_from_rs_theta})]. An alternative potential explanation is given by a small residual term in our extrapolation to the TDL; this, however, cannot conclusively be resolved based on the given data. On the other hand, we find no such systematic trend for the $\Theta$-dependence at $r_s=10$. We, thus, conclude that the present study constitutes a valuable independent cross check of current state-of-the art parametrizations of the UEG, which has further substantiated the high quality of the latter.

We are convinced that our work opens up new avenues for a gamut of future investigations. 
First, we note that the current simulation set-up might be further optimized towards the simulation of larger systems e.g.~by utilizing an improved Monte Carlo treatment of the long-range Ewald interaction~\cite{PhysRevX.13.031006} or by exploiting path contraction schemes~\cite{PhysRevE.93.043305}. Moreover, one might pursue a more efficient estimation of quantum statistics effects by combining the histogram estimator for $\mu(N,V,\beta)$ with the $\xi$-extrapolation technique~\cite{Xiong_JCP_2022,Dornheim_JCP_xi_2023,Dornheim_JPCL_2024}, which has been shown to yield highly accurate results for weak to moderate degrees of quantum degeneracy.
In addition, we note that $\mu_\textnormal{xc}(r_s,\Theta)$ constitutes a hitherto unexplored and independent alternative route towards the EOS and can, thus, be used as an additional input for the future construction of improved parametrizations that might yield superior accuracy for the evaluation of derivatives compared to current ones~\cite{status}. In a similar vein, our set-up can easily be applied to interesting parameter regimes such as the strongly coupled electron liquid regime~\cite{dornheim_electron_liquid,Tolias_JCP_2021,Tolias_JCP_2023,castello2021classical}, which are outside of the application range of either GDSMFB or corrKSDT. Moreover, we envision the application of our set-up to real WDM systems starting with light elements such as hydrogen. This will facilitate the rigorous assessment of existing EOS tables~\cite{Militzer_PRE_2021} with respect to thermodynamic consistency, and open the way towards the construction of improved parametrizations based on simulation data without the usual approximations e.g.~due to a fixed nodal structure (restricted PIMC) or an XC-functional (DFT).
Specifically, PIMC reference data for $\mu$ can be utilized to benchmark DFT results for various XC functionals. This will help to avoid problems associated with the dependence of the total free energy on the reference energy of the atoms in DFT, which can vary depending on the type of pseudopotentials used \cite{Lejaeghere_Science_2016, Kresse_1994, PhysRevB.41.7892}.

Finally, knowledge of the chemical potential opens up the \emph{Gibbs--Duhem route} to the isothermal compressibility~\cite{AttardBook}. To be more specific, an immediate consequence of the Gibbs--Duhem thermodynamic equation reads as~\cite{WidomBook,AttardBook,LombaLee1996}
\begin{equation}
\left(\frac{\partial\mu}{\partial{P}}\right)_T=\frac{1}{n}\,.
\end{equation}
Introducing the isothermal coefficient of compressibility $K_{\mathrm{T}}=-(1/V)\left(\partial{V}/\partial{P}\right)_T$, one directly obtains~\cite{AttardBook,LombaLee1996}
\begin{equation}
\frac{1}{nK_{\mathrm{T}}}=n\left(\frac{\partial\mu}{\partial{n}}\right)_T\,.
\end{equation}
Switching from the $(n,T)$ phase variables to the $(r_{\mathrm{s}},\Theta)$ phase variables, one gets
\begin{equation}
\frac{1}{nK_{\mathrm{T}}}=-\frac{r_{\mathrm{s}}}{3}\frac{\partial\mu}{\partial{r}_{\mathrm{s}}}-\frac{2\Theta}{3}\frac{\partial\mu}{\partial\Theta}\,.
\end{equation}
Within a finite difference approximation of the first order thermodynamic derivatives, the isothermal compressibility $K_{\mathrm{T}}$ can, thus, be evaluated with the histogram approach from four adjacent state point PIMC simulations. The Gibbs--Duhem route is preferable to the \emph{free energy route to the isothermal compressibility} based on~\cite{quantum_theory}
\begin{equation}
\frac{1}{nK_{\mathrm{T}}}=n\left[\frac{\partial^2}{\partial{n}^2}\left(nf\right)\right]_T\,,
\end{equation}
which, thus, also involves all possible second order derivatives with respect to $(r_{\mathrm{s}},\Theta)$, i.e. ~\cite{Tolias_PRB_2024},
\begin{align}
\frac{1}{nK_{\mathrm{T}}}&=\left(\frac{4\Theta^2}{9}\frac{\partial^2}{\partial\Theta^2}+\frac{r_{\mathrm{s}}^2}{9}\frac{\partial^2}{\partial{r}_{\mathrm{s}}^2}+\frac{4r_{\mathrm{s}}\Theta}{9}\frac{\partial^2}{\partial\Theta\partial{r}_{\mathrm{s}}}\right.\nonumber\\&\quad\,\,\,\left.-\frac{2\Theta}{9}\frac{\partial}{\partial\Theta}-\frac{2r_{\mathrm{s}}}{9}\frac{\partial}{\partial{r}_{\mathrm{s}}}\right)f\,.
\end{align}
The Gibbs-Duhem route is also preferable to the \emph{direct route to the isothermal compressibility} through the evaluation of the particle number fluctuations~\cite{hansen2013theory,BalescuBook}
\begin{equation}
nK_{\mathrm{T}}=\beta\frac{\langle{N}^2\rangle-\langle{N}\rangle^2}{\langle{N}\rangle}\,,
\end{equation}
in grand-canonical ensemble PIMC simulations, since these are generally subject to a severe fermion sign problem in WDM conditions~\cite{Dornheim_JPA_2021}. Finally, the Gibbs-Duhem route is also preferable to the \emph{sum rule route to the isothermal compressibility}, which is based on the exact long-wavelength limit of the static local field correction $G(q,0)$~\cite{quantum_theory}
\begin{equation}
\displaystyle\lim_{q\to0}G(q,0)=-\frac{q^2}{4\pi}\frac{1}{n^2K_{\mathrm{T}}^{\mathrm{XC}}}
\end{equation}
and involves the exchange-correlation contribution to the isothermal compressibility $K_{\mathrm{T}}^{\mathrm{XC}}$. The issue here is that the length scales which are relevant to the long wavelength limit are generally inaccessible in PIMC simulations due to the $N\lesssim100$ electrons simulated. However, the state-of-affairs is improving with the development of the $\xi-$PIMC extrapolation technique~\cite{Xiong_JCP_2022,Dornheim_JCP_xi_2023,Dornheim_JCP_2024}, which circumvents the fermion sign problem, thus enabling PIMC simulations with up to $\sim1000$ electrons~\cite{Dornheim_JPCL_2024} at certain parameters. Unfortunately, as aforementioned, the standard variant of the $\xi-$PIMC extrapolation technique is only applicable to mildly or moderately degenerate systems and breaks down for the UEG when $\Theta\lesssim1$~\cite{Dornheim_JCP_xi_2023,Dornheim_JCP_2024}.








\begin{acknowledgements}
This work was partially supported by the Center for Advanced Systems Understanding (CASUS), financed by Germany’s Federal Ministry of Education and Research (BMBF) and the Saxon state government out of the State budget approved by the Saxon State Parliament. This work has received funding from the European Research Council (ERC) under the European Union’s Horizon 2022 research and innovation programme
(Grant agreement No. 101076233, "PREXTREME"). Views and opinions expressed are however those of the authors only and do not necessarily reflect those of the European Union or the European Research Council Executive Agency. Neither the European Union nor the granting authority can be held responsible for them. Computations were performed on a Bull Cluster at the Center for Information Services and High-Performance Computing (ZIH) at Technische Universit\"at Dresden and at the Norddeutscher Verbund f\"ur Hoch- und H\"ochstleistungsrechnen (HLRN) under grant mvp00024.
\end{acknowledgements}

\bibliography{bibliography}
\end{document}